\documentclass[]{aa}
\usepackage{graphicx}
\usepackage{txfonts}
\usepackage{natbib,twoopt}
\usepackage[breaklinks=true]{hyperref} %% to avoid \citeads line fills
\usepackage{color}

\bibpunct{(}{)}{;}{a}{}{,}             %% natbib format for A&A and ApJ
\makeatletter
  \newcommandtwoopt{\citeads}[3][][]{\href{http://adsabs.harvard.edu/abs/#3}%
    {\def\hyper@linkstart##1##2{}%
     \let\hyper@linkend\@empty\citealp[#1][#2]{#3}}}
  \newcommandtwoopt{\citepads}[3][][]{\href{http://adsabs.harvard.edu/abs/#3}%
    {\def\hyper@linkstart##1##2{}%
     \let\hyper@linkend\@empty\citep[#1][#2]{#3}}}
  \newcommandtwoopt{\citetads}[3][][]{\href{http://adsabs.harvard.edu/abs/#3}%
    {\def\hyper@linkstart##1##2{}%
     \let\hyper@linkend\@empty\citet[#1][#2]{#3}}}
  \newcommandtwoopt{\citeyearads}[3][][]%
    {\href{http://adsabs.harvard.edu/abs/#3}
    {\def\hyper@linkstart##1##2{}%
     \let\hyper@linkend\@empty\citeyear[#1][#2]{#3}}}
\makeatother
\newcommand{\kms}{km s$^{-1}$}
\newcommand{\ci}{[C\,{\sc i}]}
\newcommand{\cii}{[C\,{\sc ii}]}
\newcommand{\oiii}{[O\,{\sc iii}]}
\newcommand{\bbarolo}{\textsc{$^{\rm 3D}$Barolo}}
\newcommand{\Msun}{M$_{\odot}$}
\newcommand{\pks}{PKS\,0529-549}
\newcommand{\sersic}{S\`ersic}

\def\llin#1 {\textcolor{blue}{#1}\;}
\def\fl#1 {\textcolor{red}{#1}\;}
\def\cdb#1 {\textcolor{orange}{#1}\;}
\definecolor{forestgreen}{rgb}{0.13, 0.55, 0.13}
\def\zy#1 {\textcolor{forestgreen}{ zy: #1}}

\begin{document} 

   \title{Gas dynamics in an AGN-host galaxy at $z\simeq2.6$:\\ regular rotation, non-circular motions, and mass models}

   \author{Lingrui Lin \inst{1,2} \and Federico Lelli\inst{3} \and Carlos De Breuck\inst{4} \and Allison Man\inst{5} \and Zhi-Yu Zhang\inst{1,2} \and Paola Santini\inst{6} \and Antonino Marasco\inst{7} \and Marco Castellano\inst{6} \and Nicole Nesvadba\inst{8} \and Thomas G. Bisbas\inst{9} \and Hao-Tse Huang\inst{10,11,5} \and Matthew Lehnert\inst{12}}

   \institute{School of Astronomy and Space Science, Nanjing University, Nanjing 210023, China
   \and Key Laboratory of Modern Astronomy and Astrophysics, Nanjing University, Ministry of Education, Nanjing 210023, China
   \and INAF - Arcetri Astrophysical Observatory, Large E. Fermi 5, 50125, Florence, Italy; \email{federico.lelli@inaf.it}
   \and European Southern Observatory, Karl-Schwarzschild-Str. 2, 85748, Garching bei München, Germany
   \and Department of Physics \& Astronomy, The University of British Columbia, 6224 Agricultural Road, Vancouver BC, V6T 1Z1, Canada
   \and INAF - Osservatorio Astronomico di Roma, Via Frascati 33, 00078, Monteporzio Catone, Italy
   \and INAF - Osservatorio Astronomico di Padova, vicolo dell’Osservatorio 5, 35122 Padova, Italy
   \and Universit\'e de la C\^ote d'Azur, Observatoire de la C\^ote d'Azur,
  CNRS, Laboratoire Lagrange, Bd de l'Observatoire, CS 34229, 06304
  Nice cedex 4, France
  \and Research Center for Astronomical Computing, Zhejiang Laboratory, Hangzhou 311100, China
  \and Department of Astronomy, University of California at Berkeley, Berkeley, CA 94720, USA
  \and Department of Physics, The Chinese University of Hong Kong, Shatin, N.T., Hong Kong
  \and Centre de Recherche Astrophysique de Lyon, ENS de Lyon, Universit\'{e} Lyon 1, CNRS, UMR5574, 69230 Saint-Genis-Laval, France}
  \titlerunning{\pks\ \ci\,(2-1) kinematics}
  \authorrunning{L. Lin et al.}
   
   \date{Received September 15, 1996; accepted March 16, 1997}

% \abstract{}{}{}{}{} 
% 5 {} token are mandatory
 
  \abstract
  {The gas dynamics of galaxies provide critical insights into the evolution of
  both baryons and dark matter (DM) across cosmic time. In this context,
  galaxies at cosmic noon --- the period characterized by the most intense star
  formation and black hole activities --- are particularly significant. In this
  work, we present an analysis of the gas dynamics of \pks: a galaxy at
  $z\simeq2.6$, hosting a radio-loud active galactic nucleus (AGN). We use new
  ALMA observations of the \ci\,(2-1) line at a spatial resolution of 0.18$''$
  ($\sim$1.5 kpc). We find that (1) the molecular gas forms a rotation-supported
  disk with $V_{\rm vrot}/\sigma_{\rm v}=6\pm3$ and displays a flat rotation
  curve out to 3.3 kpc; (2) there are several non-circular components including
  a kinematically anomalous structure near the galaxy center, a gas tail to the
  South-West, and possibly a second weaker tail to the East; (3) dynamical
  estimates of gas and stellar masses from fitting the rotation curve are
  inconsistent with photometric estimates using standard gas conversion factors
  and stellar population models, respectively; these discrepancies may be due to
  systematic uncertainties in the photometric masses, in the dynamical masses,
  or in the case a more massive radio-loud AGN-host galaxy is hidden behind the
  gas-rich \ci\ emitting starburst galaxy along the line of sight. Our work
  shows that in-depth investigations of 3D line cubes are crucial for revealing
  the complexity of gas dynamics in high-$z$ galaxies, in which regular rotation
  may coexist with non-circular motions and possibly tidal structures.}
  % % context heading (optional)
  % % {} leave it empty if necessary  
  %  {}
  % % aims heading (mandatory)
  %  {}
  % % methods heading (mandatory)
  %  {}
  % % results heading (mandatory)
  %  {}
  % % conclusions heading (optional), leave it empty if necessary 
  %  {}


   \maketitle

%
%-------------------------------------------------------------------

\section{Introduction}

The study of gas dynamics provides key insights in the formation and evolution of galaxies across cosmic time. On global scales, the distributions of baryons and dark matter (DM) shape the gravitational potential of galaxies, affecting their overall gas kinematics \citep[e.g.,][]{1986RSPTA.320..447V}. In addition, the feedback effects from massive stars (e.g., stellar winds and supernovae) and active galactic nuclei (AGN) inject energy into the interstellar medium (ISM), stirring the star-forming gas and possibly quenching the star formation of galaxies \citep[e.g.,][]{2012RAA....12..917S}. In this context, galaxies at redshift $z\simeq1-3$ are particularly important because both the cosmic star formation history and the accretion history of supermassive black holes peak around this epoch, which is known as ``cosmic noon'' \citep{2014ARA&A..52..415M}.

Rapid developments in astronomical instruments have been boosting spatially-resolved studies of gas kinematics in high-$z$ galaxies. Near-infrared (NIR) spectroscopy with integral field units (IFUs) has enabled studies of the kinematics of warm ionized gas in galaxies at $z\simeq1-3$, using H$\alpha$ or \oiii\ emission lines \citep[e.g.,][]{2009ApJ...706.1364F, 2011A&A...528A..88G, 2015ApJ...799..209W, 2019ApJ...886..124W, 2016A&A...594A..77D, 2016MNRAS.457.1888S, 2017MNRAS.471.1280T}. Radio and (sub-)millimeter interferometers can resolve kinematics of cold molecular gas, which is composed mostly of molecular hydrogen (H$_{2}$). However, the symmetric H$_{2}$ molecule does not have a permanent dipole moment so it hardly emits any lines in the cold molecular gas. Practically, we observe H$_{2}$-tracers like CO \citep{2012ApJ...760...11H,2017ApJ...841L..25T,2018MNRAS.476.3956T,2018ApJ...854L..24U,2023A&A...679A.129R,2023A&A...672A.106L}, \ci\ \citep{2018MNRAS.479.5440L,2022MNRAS.510.3734D,2022A&A...663A..22G,2023A&A...679A.129R}, or \cii\ \citep{2014A&A...565A..59D,2017ApJ...850..180J,2018Natur.553..178S,2020Natur.584..201R,2021Sci...371..713L,2021MNRAS.507.3952R} lines. 

Emission lines of different species trace distinct phases of the interstellar gas. In nearby galaxies, the warm ionized gas traced by H$\alpha$ is often found to rotate slower and has a larger velocity dispersion than the cold molecular gas traced by low-$J$ CO lines \citep[e.g.,][]{2018ApJ...860...92L, 2022ApJ...934..173S}. In galaxies hosting starbursts and/or AGNs, emission lines of H$\alpha$ and \oiii\ can be dominated by galactic outflows \citep{2014A&A...568A..14A,2014MNRAS.441.3306H,2017A&A...606A..36C,2019A&A...622A.188C,2022MNRAS.513.2535C}, which further complicates analyses of galaxy rotation.

For high-$z$ galaxies, the beam smearing effect \citep{1973MNRAS.163..163W,1978PhDT.......195B,1989A&A...223...47B} also becomes significant as usually these galaxies can be spatially resolved only with a few independent elements with current facilities. This will lead to the observed emission lines being broadened by both the intrinsic turbulent motion of the interstellar gas and the unresolved rotation velocity ($V_{\rm rot}$) structure within the telescope beam. In addition, the observed line-of-sight velocity is intensity-weighted so it is biased towards small galactic radii where the surface brightness is higher. To overcome the beam smearing effect, various tools have been developed to fit a rotating disk model directly to the 3-dimensional (3D) emission-line cubes \citep[e.g.,][]{2015AJ....150...92B,2015MNRAS.451.3021D,2015MNRAS.452.3139K}.

Therefore, to study the dynamics of high-$z$ galaxies, especially the extreme cases at cosmic noon, one needs multi-phase gas tracers for a panchromatic view of both circular and non-circular motions as well as careful treatment of beam smearing effects (i.e., high-resolution data and reliable modeling). To date, such studies are only limited to a handful of cases \citep[e.g.,][]{2017ApJ...846..108C, 2018ApJ...854L..24U, 2018MNRAS.479.5440L, 2023A&A...672A.106L}.

\pks\ is a well-studied radio galaxy at $z\simeq2.57$ with plenty of multi-wavelength data --- optical spectroscopy, NIR imaging, and radio polarimetry \citep{2007MNRAS.375.1059B}, Spitzer Infrared Array Camera (IRAC), Infrared Spectrograph (IRS), and Multiband Imaging Photometer for Spitzer (MIPS) imaging \citep{2010ApJ...725...36D}, Herschel Photodetector Array Camera and Spectrometer (PACS) and Spectral and Photometric Imaging Receiver (SPIRE) imaging \citep{2014A&A...566A..53D}, 1.1-mm data from AzTEC \citep{2011MNRAS.418...74H}, Very Large Telescope (VLT) Spectrograph for INtegral Field Observations in the Near Infrared (SINFONI) imaging spectroscopy \citep{2017A&A...599A.123N}, Atacama Large Millimeter Array (ALMA) \ci\,(2-1) line \citep{2018MNRAS.479.5440L} and band-6 continuum \citep{2019A&A...621A..27F}, and VLT/X-Shooter spectra from rest-frame ultra-violet (UV) to optical \citep{2019A&A...624A..81M}.

\pks\ hosts a Type-{\sc ii} AGN and two radio lobes \citep{2007MNRAS.375.1059B}. The Eastern lobe has the highest Faraday rotation measure ever observed to date \citep{2007MNRAS.375.1059B}, suggesting that the galaxy is surrounded by a medium with high electron density and/or a strong magnetic field. \pks\ has an estimated stellar mass ($M_{\star}$) of $3\times10^{11}$ \Msun\ \citep{2010ApJ...725...36D} derived by fitting the stellar spectral energy distribution (SED), and a star formation rate (SFR) of $1020_{-170}^{+190}$ \Msun\ yr$^{-1}$ \citep{2019A&A...621A..27F} derived by the total infrared luminosity. \pks\ has experienced at least two bursts of recent star formation in the past, 6 Myr and $>$ 20 Myr, respectively, based on an analysis of the photospheric absorption features in the rest-frame UV spectrum \citep{2019A&A...624A..81M}.

Using ALMA observations, \citet{2018MNRAS.479.5440L} found that the \ci\,$^{3}$P$_{2}\rightarrow^{3}$P$_{1}$ emission (hereafter \ci\,(2-1)) is consistent with a rotating disk. The \oiii\,$\lambda5007$ emission \citep{2017A&A...599A.123N}, on the other hand, is more extended and is aligned with the radio lobes, so it is probably dominated by an AGN-driven outflow. The rotation speed of the gas disk traced by \ci\ provided a total dynamical mass consistent with the observed baryonic mass, but detailed mass models that separate the gravitational contributions of baryons and/or DM could not be constructed due to the low resolution and sensitivity of the \ci\,(2-1) data. For the same reasons, it was not possible to measure the gas velocity dispersion and to investigate possible non-circular motions in the molecular disk.

In this work, we present new ALMA \ci\,(2-1) observations of \pks\ with high spatial resolution and sensitivity. The \ci\ lines are among the most efficient H$_{2}$-tracers for galaxies at cosmic noon because they are accessible through ALMA band 4 and band 6. At $z\sim2$, the CO lines ($J\geq3$) covered by ALMA are weak. The \cii\,158-$\mu$m line, instead, is difficult to observe at $z\simeq1-3$ due to its high frequency (even though redshifted) that requires excellent weather conditions at the ALMA site, but it is cost-effective for galaxies at $z>4$ because it becomes observable with ALMA band 7 \citep[e.g.,][]{2014A&A...565A..59D,2017ApJ...850..180J,2018Natur.553..178S,2021Sci...371..713L}.

This paper is structured as follows. Section~\ref{sec: data} describes the new ALMA observations and the data reduction. Section~\ref{sec:gasdust} describes the gas and dust distribution as well as their radial surface brightness profile. Section~\ref{sec: rotation curve fitting} studies the gas kinematics and measures the rotation curve of \pks\ as well as non-circular motions. Section~\ref{sec: mass models} builds mass models with different combinations of baryonic and DM components, testing the consistency of our observations with the expectations from the $\Lambda$ cold dark matter ($\Lambda$CDM) cosmology. Section~\ref{sec: discussion} discusses the implications of our results. Section~\ref{sec: conclusions} provides a summary.

Throughout this paper, we assume a flat $\Lambda$CDM cosmology with $H_{0}=$ 67.4 \kms\ Mpc$^{-1}$, $\Omega_{\rm m}$ = 0.315, and $\Omega_{\Lambda}$ = 0.685 \citep{2020A&A...641A...6P}. In this cosmology, 1 arcsec corresponds to 8.22 kpc at the redshift of \pks\ ($z=2.57$), while the age of the Universe and the lookback time are 2.5 Gyr and 11.3 Gyr, respectively.

%--------------------------------------------------------------------
\section{Data analysis}
\label{sec: data}

\subsection{ALMA observations}

The ALMA band-6 observations were carried out during ALMA Cycle 6 (Project ID: 2018.1.01669.S, PI: Federico Lelli), targeting the \ci\,(2-1) line. Four spectral windows were centered at 226.200, 228.075, 240.000, and 241.875 GHz --- each covers 1.875 GHz with 480 channels for a native velocity resolution of 5 \kms. The first spectral window was chosen to cover both the \ci\,(2-1) line (rest-frequency of 809.341970 GHz) and the CO\,$J=7-6$ line (rest-frequency of 806.651806 GHz); the other three spectral windows cover the continuum emission. Three execution blocks (EBs) were conducted on 9 August, 23 August, and 18 September, respectively, in 2019. The on-source times were 32.93, 32.93, and 15.20 min, respectively (1.35 hours in total). The first EB was labeled as semi-pass in the initial quality assurance (QA0) but we kept this EB because it improved the imaging quality after careful manual calibration. The latter two EBs (QA2 pass) were calibrated using the standard Common Astronomy Software Applications (CASA) pipeline (v5.6.1-8) \citep{2022PASP..134k4501C}.

\subsection{Imaging and cleaning}

Imaging of the \ci\,(2-1) cube was interactively done with the \texttt{tclean} task in CASA (v6.5.2.26), using Briggs' weighting with a robust parameter of 1.5 and a uv-taper of 0.05 arcsec. This gave a restored beam of 0.178 arcsec $\times$ 0.163 arcsec with a position angle $\mathrm{PA}=-20.9$ deg. To reach an optimal compromise between resolution and sensitivity, we circularized the beam to 0.18 arcsec and rebinned the velocity resolution to 25.8 \kms. The root-mean-square (RMS) noise of the final \ci\,(2-1) cube is $\sim0.15$ mJy beam$^{-1}$. The continuum of the \ci\,(2-1) cube was subtracted by fitting a zeroth-order polynomial using the line-free channels (227.4047 to 228.9916 GHz) in the image plane. The CO\,$J=7-6$ line in the same spectral window was masked by visually trimming the spectrum.

A band-6 continuum image was created by combining the spectral windows centered at 228.075, 240.000, and 241.875 GHz. We used \texttt{tclean} in interactive mode with Briggs' weighting, robust parameter of 1.5, and a uv-taper of 0.1 arcsec. This gave a restored beam of 0.199 arcsec $\times$ 0.195 arcsec with a position angle $\mathrm{PA}=89.5$ deg. The RMS noise of the band-6 continuum image is $\sim0.012$ mJy beam$^{-1}$.

\section{Dust and gas distribution} \label{sec:gasdust}

\subsection{Global overview}

\begin{figure}[h!]
\centering
\includegraphics[width=\hsize]{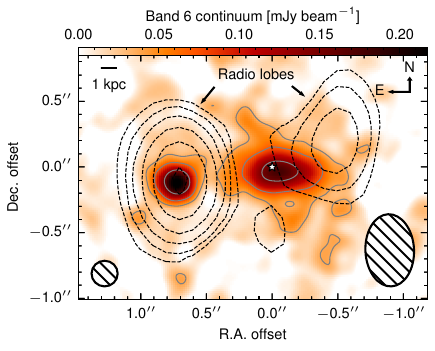}
   \caption{The ALMA band-6 continuum map (background color map and grey solid contours) overlaid with the ATCA 18-GHz continuum map (black dashed contours). The axis coordinates are relative to the kinematic center (white star). The synthesized beams of the ALMA and the ATCA observations are shown in the bottom-left and the bottom-right corners, respectively. Contours in the ALMA map are at $S/N=3\times(1, 2, 4)$, while those in the ATCA map are at $S/N=3\times(1, 2, 4, 8, 16, 32)$.}
\label{fig: continuum-maps}
\end{figure}

\begin{figure*}[h!]
\centering
\includegraphics[width=\hsize]{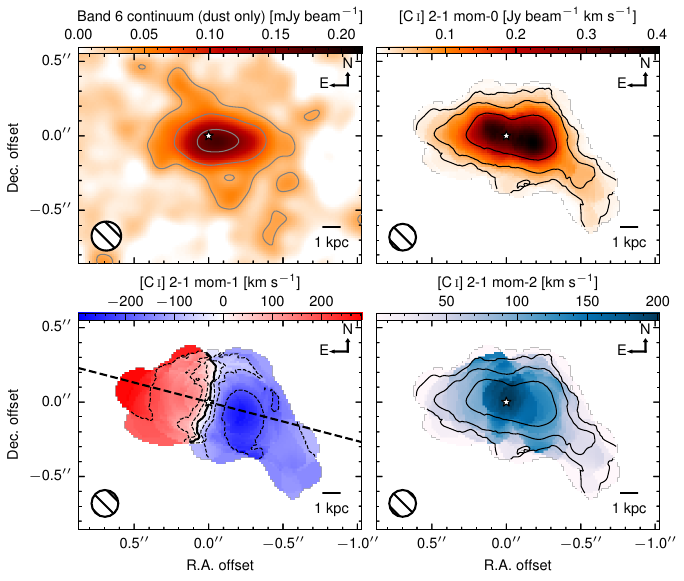}
   \caption{Maps of dust-only continuum emission (top left), \ci\,(2-1) integrated intensity (moment-0, top right), intensity-weighted velocity (moment-1, bottom left), and intensity-weighted line broadening (moment-2, bottom right). The axis coordinates are relative to the kinematic center (white star). The synthesized beams are shown in the lower-left corner of each panel. In the dust-only continuum map, contours correspond to signal-to-noise ratios $S/N=3\times(1, 2, 4)$. The RMS noise of the dust continuum is $\sim0.012$ mJy beam$^{-1}$. In the moment-0 map, the contours correspond to $S/N=3\times(1, 2, 4)$. The $S/N$ is calculated by \bbarolo, which follows the procedure in \citet{Lelli2014}. In the moment-1 map, the black dashed contours show line-of-sight velocity of $\pm50,\pm100$, and $\pm200$ \kms\ with respect to the systemic velocity (set to zero, black solid contour). The black dashed line represents the kinematic major axis. In the moment-2 map, the contours are the same as in the moment-0 map.}
\label{fig: moment-maps}
\end{figure*}

Fig.~\ref{fig: continuum-maps} shows the ALMA band-6 (354 $\mu$m at rest-frame of \pks) continuum map at $\sim0.2''$ resolution. We confirm the two continuum components discussed in \citet{2018MNRAS.479.5440L}: the one on the East coincides with the radio lobe from the Australia Telescope Compact Array (ATCA) 18-GHz observations, while the one in the middle coincides with the \ci\ (Fig.~\ref{fig: moment-maps}) and optical emission. \citet{2019A&A...621A..27F}, indeed, have shown that the Eastern compact source is consistent with the extrapolation of the power-law synchrotron emission from the radio band to the sub-mm band, while the central emission coincides with the gas disk and is contributed mostly by the cold dust in the galaxy. The new ALMA data confirm this interpretation.

To create a continuum map that contains only dust emission, we used the following procedure to remove the Eastern synchrotron component. We first interactively drew a mask enclosing only the synchrotron component in the \texttt{tclean} task (with \texttt{savemodel=`modelcolumn'}). Next, we subtracted the synchrotron clean component from the original uv-data (using the \texttt{uvsub} task) and redid the imaging. The top-left panel of Fig.~\ref{fig: moment-maps} shows the outcome of this procedure.

The \ci\,(2-1) moment maps were constructed with 3D-Based Analysis of Rotating Object via Line Observations \citep[\bbarolo,][]{2015MNRAS.451.3021D}, considering the \ci\,(2-1) signal within a fiducial mask created by the task \texttt{Smooth \& Search} (see Section~\ref{sec: rotation curve fitting}). The moment maps are shown in Fig.~\ref{fig: moment-maps}. The new ALMA data unambiguously confirm and spatially resolve the rotating disk identified in \citet{2018MNRAS.479.5440L}. In addition, the moment maps reveal a gas tail to the South-West of the rotating disk. In Section~\ref{sec: non-circular}, we present a detailed 3D analysis which reveals a kinematically anomalous component in the blueshifted side of the disk and a second, weaker gas tail in the redshifted side to the East. It is possible that these three different non-circular components have a common physical origin, as we discuss in Section~\ref{sec: discuss non-circular motions}.

\subsection{Radial surface brightness profiles}
\label{sec: densityprofile}

Radial surface brightness profiles provide an effective 1-D description of the dust and gas distribution in galaxies. They are useful for measuring characteristic scale lengths, such as the effective radius ($R_{\rm e}$) that contains half of the total flux. They are also needed to study the overall mass distribution of galaxies because one needs radial surface density profiles of different baryonic components to compute their gravitational field in the galaxy mid-plane out to infinity (see section~\ref{sec: mass models}).

\begin{figure}[h!]
\centering
\includegraphics[width=\hsize]{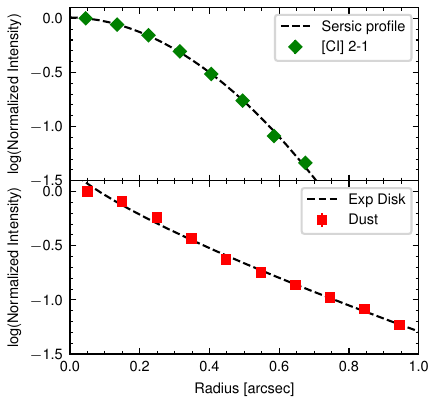}
   \caption{Radial surface brightness profiles of \ci\,(2-1) (top, green diamonds) and dust continuum (bottom, red squares). The random error of each data point is $\lesssim6\%$. The black dashed lines show \sersic\ fits to each profile.}
   \label{fig: densityprofile}
\end{figure}

For the dust radial profile, we use the synchrotron-subtracted band-6 continuum image (top-left panel in Fig.~\ref{fig: moment-maps}). We measure the surface brightness profile averaging over a set of elliptical annuli, positioned according to the kinematic center, the inclination and position angle of the rotating disk (as derived in Section~\ref{sec: disk geometry}). The annulus spacing is half of the beam size of the dust continuum (0.10 arcsec). The results are shown in Fig.~\ref{fig: densityprofile} (top panel).

UV emission from both young stellar populations and the AGN can heat up the dust grains, though the latter is expected to contribute a negligible fraction to the Rayleigh-Jeans tail of the cold dust emission \citep{2019A&A...621A..27F,2021A&A...654A..90L}. To test this effect, we measure the radial profile using also the ALMA band-4 continuum map at 625 $\mu$m rest-frame \citep{2024arXiv241104290H}. We find that the dust emission profile at band 4 is consistent with that at band 6, confirming that the dust continuum at band 6 is dominated by cold dust grains heated by young stars. Actually, \citet{2019A&A...624A..81M} have shown that the UV emission from the host galaxy is dominated by the young stellar population, not the AGN.

For the gas radial profile, we use a \ci\,(2-1) moment-0 map integrated within $(-550,550)$ \kms\ centered at the \ci\,(2-1) systematic velocity (Section~\ref{sec: rotation curve fitting}) to account for possible faint emissions missed by the source mask. We measure the surface brightness profile using the same set of annuli as for the dust profile. The annuli spacing is half of the beam size of the \ci\,(2-1) cube (0.09 arcsec). The results are shown in Fig.~\ref{fig: densityprofile} (bottom panel).

Both profiles are fitted with a \sersic\ function \citep{1968adga.book.....S} parameterized by the \sersic\ index ($n$) and the effective radius ($R_{\rm e}$). The fitting is done using the orthogonal-distance-regression method in \texttt{scipy.odr}. For the dust profile, we fit the data out to radii $R=1$ arcsec and obtain $n=1.3\pm0.2$. For the gas profile, we fit only the data at $R<0.7$ arcsec because we aim to trace the inner rotating disk without contribution from the outer gas tail. We obtain $n=0.52\pm0.01$, which properly captures the inner flattening of the gas profile. The best-fit radial density profiles are shown in Fig.~\ref{fig: densityprofile} and the best-fit parameters are given in Table~\ref{table: density profile}.

\begin{table}[h!]
\caption{Outputs of \sersic\ fits to the gas and dust surface brightness profiles.}     % title of Table
\label{table: density profile}      % is used to refer this table in the text
\centering    % used for centering table
\footnotesize
\renewcommand{\arraystretch}{1.2}
\begin{tabular}{cccc}        % centered columns (2 columns)
\hline\hline                 % inserts double horizontal lines
Component     & \sersic\ $n$    & $R_{\rm e}$ (arcsec)  & $R_{\rm e}$ (kpc)\\    % table heading
\hline                        % inserts single horizontal line
Gas   & $0.52\pm0.01$ & $0.313\pm0.003$ & $2.57\pm0.02$\\
Dust  & $1.3\pm0.2$   & $0.57\pm0.04$  & $4.7\pm0.3$ \\
\hline                                   %inserts single line
\end{tabular}
\end{table}

\section{Gas kinematics}
\label{sec: rotation curve fitting}

We study the gas kinematics of \pks\ using \bbarolo\ \citep{2015MNRAS.451.3021D}. In \bbarolo, a rotating disk is modeled with a set of tilted rings, each characterized by five geometric parameters --- center coordinates ($x_{0}$, $y_{0}$), systemic velocity ($V_{\rm sys}$), position angle ($PA$), and inclination ($i$) --- and five physical parameters --- rotation velocity ($V_{\rm rot}$), radial velocity ($V_{\rm rad}$), velocity dispersion ($\sigma_{\rm v}$), surface density ($\Sigma_{\rm gas}$), and vertical thickness ($z_{0}$). The tilted-ring model is convolved with the telescope beam and then is iteratively compared with the observations to obtain the best-fit parameters.

The 3D fit of \bbarolo\ is performed on a masked cube which includes mostly real line emission and avoids noisy pixels. We generate the source mask by setting \texttt{MASK=SMOOTH\&SEARCH}, \texttt{FACTOR=1.8} (factor by which the cube is spatially smoothed before source search), \texttt{SNRCUT}=4 (primary S/N threshold), \texttt{GROWTHCUT=3} (secondary S/N threshold to growth the primary mask), and \texttt{MINCHANNELS=2} (minimum number of channels for an accepted detection). A different choice of the source mask would not substantially change our general results.

Within the source mask, the \ci\,(2-1) disk of \pks\ can be fitted with five rings. The width of each ring is 0.09 arcsec, which is half of the beam size of the \ci\,(2-1) cube. We set \texttt{NORM=AZIM} so that the observed moment-0 map is azimuthally averaged to obtain the $\Sigma_{\rm gas}$ of each ring in the model. For the vertical density distribution, we assume an exponential profile (\texttt{LTYPE=3}) with a fixed scale height of 300 pc ($\sim$ 0.04 arcsec). The disk scale height is much smaller than the \ci\,(2-1) beam (0.18 arcsec) so it has negligible impact on the kinematic fitting. We also fix $V_{\rm rad}=0$ because there are no indications for strong radial motions, which generally produce a non-orthogonality between the kinematic major and minor axes \citep[e.g.,][]{Lelli2012a, Lelli2012b, DiTeodoro2021}. Therefore, seven free parameters need to be optimized: $x_0$, $y_0$, $V_{\rm sys}$, $PA$, $i$, $V_{\rm rot}$, and $\sigma_{\rm v}$. To obtain the rotation curve, we first estimate the geometric parameters and then fit the kinematic parameters ($V_{\rm rot}$ and $\sigma_{\rm v}$) with the disk geometry fixed.

\subsection{Disk geometry}
\label{sec: disk geometry}

We first run \texttt{3DFIT} on the \ci\,(2-1) cube, leaving all seven parameters free. To estimate the overall geometry, all pixels are uniformly weighted (\texttt{WFUNC=0}) and both sides of the rotating disk are considered (\texttt{SIDE=B}). We set the initial $PA=75^{\circ}$ and initial $i=50^{\circ}$. We also set \texttt{DELTAPA=15} and \texttt{DELTAINC=15} such that $PA$ and $i$ can explore the parameter space within $\pm15^{\circ}$ around their initial guesses. The initial value of $\sigma_{\rm v}$ is 30 \kms \citep{2018MNRAS.479.5440L}. We let \bbarolo\ guess the initial values of $V_{\rm sys}$ and $V_{\rm rot}$ automatically.

After several tests, we find that the kinematic center is difficult to measure because the best-fit value does not coincide with the kinematic minor axis (defined by the iso-velocity contour equal to $V_{\rm sys}$) as expected for a rotating disk. This is likely due to the disturbed gas kinematics on the approaching side of the disk. Therefore, we fix the kinematic center of the galaxy to (R.A., Dec.) = ($\rm 5^{h}30^{m}25.447^{s}$, $-54^{\circ}54'23.165''$) so that it lies along the kinematic minor axis (see the bottom left panel of Fig.~\ref{fig: moment-maps}), and re-run the fits with five free parameters.

Table \ref{table: disk geometry} summarizes the disk geometric parameters fitted by \bbarolo. The adopted values of $V_{\rm sys}$, $PA$, and $i$ are measured as the median values across different rings. The uncertainties are estimated as
\begin{equation}\label{eq:errors}
    \delta = \sqrt{ \dfrac{1}{\rm N} \rm{MAD}^2 + \dfrac{1}{\rm N^{2}}\sum^{N}_{i} \delta_{i}^2 }
\end{equation}
where $N=5$ is the number of rings, MAD is the median absolute deviation across the rings, and $\delta_{i}$ are the individual errors on the given parameter at each ring. Under the radical sign, the first term considers the variation among different rings while the second term considers the uncertainty of each ring.

The best-fit $PA$ and $i$ are perfectly consistent with the values from \citet{2018MNRAS.479.5440L} of $75^{\circ}$ and $50^{\circ}$, respectively. For such an inclination angle, $V_{\rm rot}$ is not sensitivity to the inclination correction; for example, $V_{\rm rot}$ only changes by $\sim$10$\%$ when $i$ varies from $50^{\circ}$ to $60^{\circ}$. 

\begin{table}[h!]
\caption{Disk geometric parameters of \pks.}             % title of Table
\label{table: disk geometry}      % is used to refer this table in the text
\centering                          % used for centering table
\footnotesize
\renewcommand{\arraystretch}{1.2}
\begin{tabular}{c c}        % centered columns (2 columns)
\hline\hline                 % inserts double horizontal lines
Parameter & Value  \\    % table heading 
\hline                        % inserts single horizontal line
R.A. (J2000)           & $\rm 5^{h}30^{m}25.447^{s}$ \\
Dec. (J2000)           & $-54^{\circ}54'23.165''$ \\
$V_{\rm sys}$ (\kms)   & $47\pm16$ \\
$PA$ ($^{\circ}$)      & $75\pm7$ \\
$i$ ($^{\circ}$)       & $53\pm5$  \\
\hline                                  %inserts single line
\end{tabular}
\tablefoot{$V_{\rm sys}$ is relative to $z=2.570$, as measured by the \ci\,(2-1) emission from \citet{2018MNRAS.479.5440L}. Uncertainties are calculated using Eq.\,\ref{eq:errors}.} 
\end{table}

\subsection{Rotation velocity and velocity dispersion}
\label{sec: Vrot and sigma_v}

Fixing the geometric parameters, we run \texttt{SPACEPAR} in \bbarolo\ to look for global minima in the parameter space of $V_{\rm rot}-\sigma_{\rm v}$. We explore $V_{\rm rot}$ within [200, 450] \kms\ and $\sigma_{\rm v}$ within [1, 200] \kms, both with a grid step of 1 \kms. The residual function to be minimized is $|M-D|$ (\texttt{FUNC=2}), where $M$ and $D$ are the intensity values at each 3D voxel of the model and the data cube, respectively. To examine the effect of non-circular motions (such as the enhanced kinematic irregularities on the blueshifted side), we run \texttt{SPACEPAR} separately on the approaching (blueshifted, \texttt{SIDE=A}) and receding sides (redshifted, \texttt{SIDE=R}), as well as simultaneously on both sides (\texttt{SIDE=B}).

\begin{figure}
\centering
\includegraphics[width=\hsize]{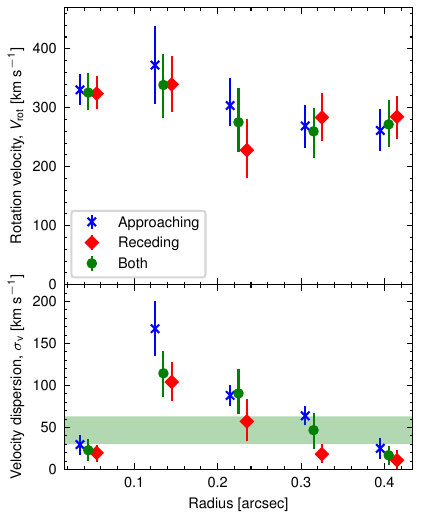}
   \caption{Rotation velocity and velocity dispersion of the \ci\,(2-1) disk by fitting the approaching side (blue crosses), the receding side (red diamonds), and both sides (green dots). The green band is centered at the median $\sigma_{\rm v}=47$ \kms\ from the two-sides fitting and has a half-width of 16 \kms, representing the uncertainty of $\sigma_{\rm v}$ (see Section~\ref{sec: Vrot and sigma_v}).}
   \label{fig: Vrot-Vdisp-Spacepar}
\end{figure}

\begin{figure}[h!]
\centering
\includegraphics[width=\hsize]{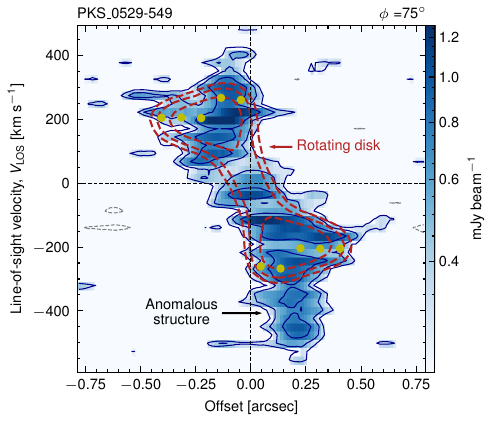}
   \caption{Position-velocity diagram along the kinematic major axis. The systemic velocity of the rotating disk is set at $V_{\rm LOS}=0$ \kms. The color-scale and the blue contours show the observed \ci\,(2-1) data. The red contours and yellow dots show, respectively, the best-fit rotating disk model and the rotation velocity projected along the line of sight, assuming a constant velocity dispersion ($\sigma_{\rm v}=47$ \kms). The contour levels are at $S/N=(\pm2,3,5)$; negative contours are shown with gray dashed lines.}
   \label{fig: ci-pv}
\end{figure}

Fig.~\ref{fig: Vrot-Vdisp-Spacepar} shows $V_{\rm rot}$ and $\sigma_{\rm v}$ of each ring optimized on different sides. The rotation velocities are consistent within the errors among the three different runs. When fitting only the approaching side, the velocity dispersion shows an elevated value at $R\simeq0.135''$, which is likely due to complex non-circular motions rather than a real increase in the gas turbulence (see Fig.~\ref{fig: ci-pv}). The non-circular motions are examined in detail in Section~\ref{sec: non-circular}.

The current data are unable to properly constrain the radial profile of the gas velocity dispersion, so we calculate the median $\sigma_{\rm v}$ from the two-sides fitting (47$\pm$16 \kms) and use it as our fiducial estimate of the intrinsic gas velocity dispersion. The uncertainty is calculated using Eq.\,\ref{eq:errors}. This measurement of $\sigma_{\rm V}$ is consistent within the errors with the fiducial upper limit of $\sim$30 \kms\ estimated by \citet{2018MNRAS.479.5440L}.

As a final step, we rerun \texttt{3DFIT}, fixing $\sigma_{\rm v}=47$ \kms\ and leaving only $V_{\rm rot}$ free. We set \texttt{WFUNC=2} to give more weights along the kinematic major axis. Fig.~\ref{fig: ci-pv} compares the position-velocity (PV) diagram along the kinematic major axis of the observed cube with the best-fit model cube. Overall, the disk model provides a good description of the observations. In particular, the thickness of the observed PV diagram is well reproduced by the model, indicating that the velocity dispersion is reasonable. Non-circular motions that cannot be reproduced by the rotating disk model will be described in detail in Section~\ref{sec: non-circular}.

\subsection{Asymmetric drift correction}

The gas disk of \pks\ is rotationally supported, having a median $V_{\rm rot}/\sigma_{\rm v}=6\pm3$. The uncertainty is calculated by propagating the errors on $V_{\rm rot}$ and $\sigma_{\rm v}$, which are estimated using Eq.~\ref{eq:errors}. Turbulent motions, however, may provide non-negligible pressure support, so we estimate the asymmetric drift correction (ADC) to obtain the circular velocity ($V_{\rm c}$) that directly relates to the gravitational potential.

\begin{figure}[h!]
\centering
\includegraphics[width=\hsize]{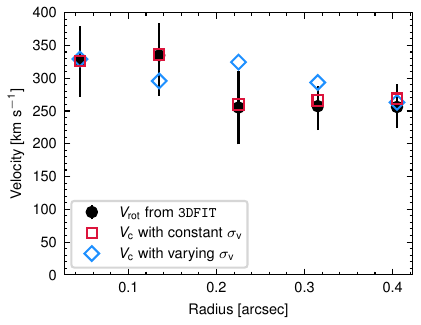}
   \caption{The circular-velocity curve of \pks\ after correcting for pressure support. The black dots with errorbars show the observed rotation velocity from \texttt{3DFIT}. The red squares and the blue diamonds show the circular velocities after asymmetric drift correction assuming a constant $\sigma_{\rm v}$ and a radially varying $\sigma_{\rm v}$, respectively.} 
   \label{fig: ADC}
\end{figure}

The ADC depends on the radial gradients of $\sigma_{\rm v}$ and $\Sigma_{\rm gas}$ \citep[see, e.g., Eq. 4 in][]{Lelli2023}. In \bbarolo, the ADC can be computed using polynomials to describe the radial profiles of $\sigma_{\rm v}$ and $\Sigma_{\rm gas}$. Fig.\,\ref{fig: ADC} shows the resulting $V_{\rm c}(R)$ assuming a constant $\sigma_{\rm v}=47$ \kms\ or a radially varying $\sigma_{\rm v}(R)$ (taken from the two-sides fitting). We find that the values of $V_{\rm rot}$ and the two versions of $V_{\rm c}$ are consistent within the errors, confirming that the rotation support is dominant while pressure support is nearly negligible. Hereafter, we use $V_{\rm c}$ from a radially constant $\sigma_{\rm v}$ for simplicity.

\subsection{Non-circular motions}
\label{sec: non-circular}

\begin{figure*}
\centering
\includegraphics[width=\textwidth]{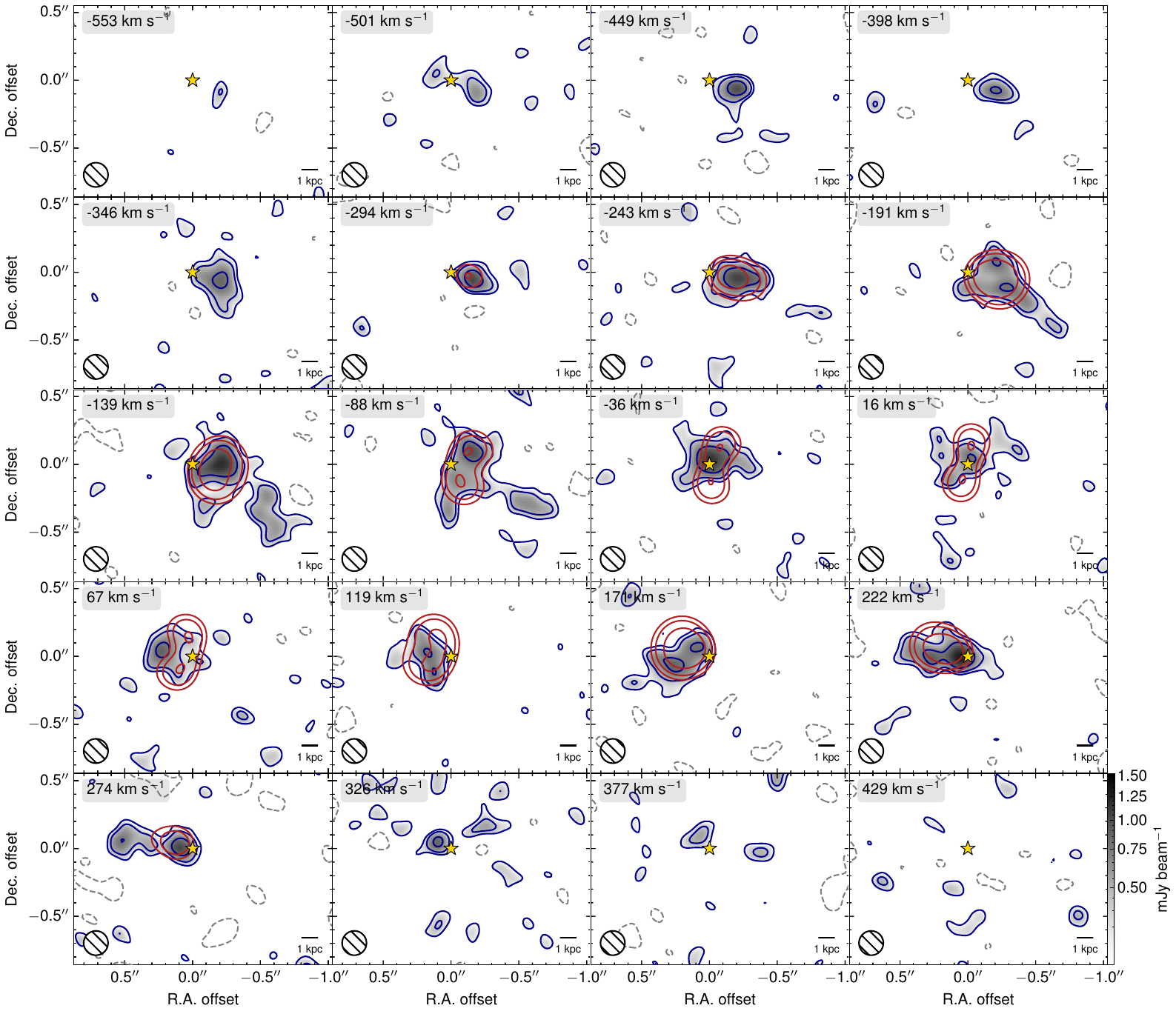}
   \caption{Channel maps of \ci\,(2-1) cube, showing every two channels. The grayscale and the blue contours show the observed \ci\,(2-1) data. The red contours show the best-fit rotating disk model (Section.~\ref{sec: rotation curve fitting}). The contour levels are at $S/N=(\pm 2,3,5)$; negative contours are shown with gray dashed lines. The axis coordinates are relative to the kinematic center (golden star). The line-of-sight velocities are relative to the redshift of \ci\,(2-1) line. The synthesized beam of the \ci\,(2-1) cube is shown in the bottom-left corner of each panel.}
   \label{fig: channel-map}
\end{figure*}

\begin{figure*}
  \resizebox{\hsize}{!}
  {
  \includegraphics[width=0.51\textwidth]{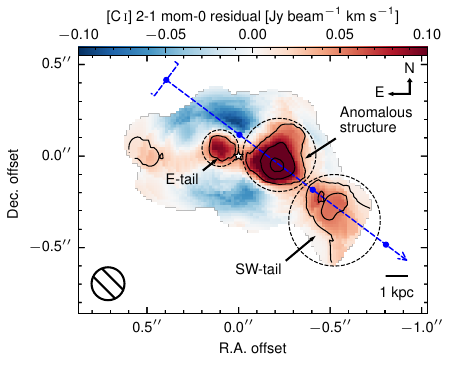}
  \includegraphics[width=0.49\textwidth]{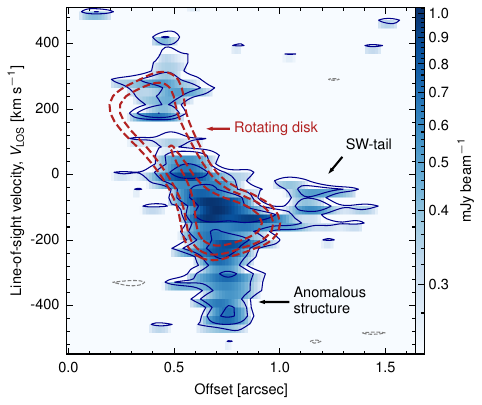}
  }
   \caption{Left: Residual \ci\,(2-1) map obtained by subtracting the best-fit rotating disk model from the observed moment-0 map. The synthesized beam of \ci\ data is shown at the lower-left. The axis coordinates are relative to the kinematic center (white star). The dashed circles show the aperture to measure the flux of the enclosed structures. The black contours correspond to $S/N=3\times(1, 2, 3)$ of the residual map. The blue dashed arrow indicates the path for extracting the PV diagram on the right. The upper-left end of the arrow indicates the width of the path. The blue dotted tickmarks indicate offsets of 0.0 arcsec, 0.5 arcsec, 1.0 arcsec, and 1.5 arcsec along the path. Right: Position-velocity diagram along the path on the left. The color coding and the contour levels are the same as those in Fig.~\ref{fig: ci-pv}.}
   \label{fig: ci-mom0-residual}
\end{figure*}

The channel maps (Fig.~\ref{fig: channel-map}) show that there are non-circular motions that cannot be reproduced by the rotating disk model: 1) a gas tail to the South-West of the rotating disk at line-of-sight (LoS) velocities from $-191$ to $-88$ \kms\ (SW-tail); 2) a second weaker gas tail to the East of the rotating disk at LoS velocities from $274$ to $377$ \kms (E-tail); and 3) an anomalous structure at $R\simeq0.1-0.3''$ at LoS velocities from $-501$ to $-346$ \kms\ (see also Fig.~\ref{fig: ci-pv}). These non-circular components are also visible in the residual \ci\,(2-1) moment-0 map (left panel of Fig.~\ref{fig: ci-mom0-residual}), which is obtained by subtracting the best-fit model cube from the observed cube.

\begin{figure}[h!]
\centering
\includegraphics[width=\hsize]{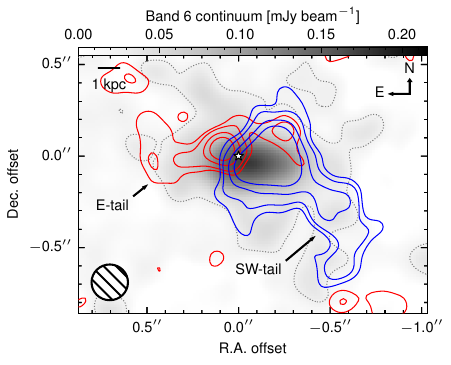}
   \caption{\ci\,(2-1) Renzograms (red and blue contours) overlaid on the dust-only continuum map (grayscale and gray dotted contours). The blue and red contours show ``narrow'' \ci\,(2-1) moment-0 maps integrated within LoS velocities from $-191$ to $-88$ \kms\ and $274$ to $377$ \kms, respectively. The contour levels are at $S/N=(2,3,5)$. A single contour at $S/N=2$ of the dust continuum is also shown. The synthesized beam of the dust continuum is shown in the lower-left corner.}
   \label{fig: gas tails}
\end{figure}

To better visualize the tail-like structures, we construct the so-called ``Renzograms'' \citep{1976A&A....53..159S} from the \ci\,(2-1) cube by integrating over the velocity intervals of the two gas tails specified above, and overlay them on the dust-only continuum map (Fig.~\ref{fig: gas tails}). While the contours around the kinematic center are influenced by emission from the rotating disk (especially for the blue contours), the outer parts trace mostly the gas tails, possibly extending beyond the main body of the galaxy disk. The SW-tail is significantly detected at $S/N>3$ while the E-tail is detected at $S/N\sim2-3$.

The right panel of Fig.~\ref{fig: ci-mom0-residual} shows a PV diagram extracted along the path in the left panel, averaging over a width of 0.225 arcsec. It is difficult to tell whether the anomalous structure and the SW-tail are kinematically connected because the eventual connection occurs at the same LoS velocities of the gas disk. The physical nature of these three non-circular components remains unclear and will be discussed in Section~\ref{sec: discuss non-circular motions}. A sensible hypothesis is that we are seeing two "leftover" tidal tails due to a past major merger and a gas inflow towards the galaxy center, possibly related to the SW-tail.

Using the residual \ci\,(2-1) moment-0 map, we estimate the \ci\,(2-1) flux associated with the non-circular motions. We sum over pixels with $S/N>3$ enclosed by the apertures shown in the left panel of Fig.~\ref{fig: ci-mom0-residual}. The fluxes and the fiducial uncertainties are given in Table~\ref{table: non-circular flux} but we stress that these values are lower limits. In fact, given that the rotating disk model assumes axis-symmetry, the flux in the observed moment-0 map is azimuthally averaged over rings, including the flux of the non-circular motions. This explains why the residuals along the minor axis of the galaxy are systematically negative. The low-$S/N$ pixels associated with the gas tails are also discounted. The non-circular motions are responsible for at least $12\%$ of the total flux in the gas disk ($2.8\pm0.3$ Jy \kms). We will further discuss the non-circular motions in Section~\ref{sec: discuss non-circular motions}.

\begin{table}
\caption{\ci\,(2-1) fluxes of the non-circular components.}             % title of Table
\label{table: non-circular flux}      % is used to refer this table in the text
\centering                          % used for centering table
\footnotesize
\renewcommand{\arraystretch}{1.2}
\begin{tabular}{c c}        % centered columns (2 columns)
\hline\hline                 % inserts double horizontal lines
Structure & Flux (Jy \kms)  \\    % table heading 
\hline                        % inserts single horizontal line
Anomalous structure & $0.20\pm0.02$ \\
SW-tail             & $0.10\pm0.01$  \\
E-tail              & $0.030\pm0.003$ \\
\hline                                  %inserts single line
\end{tabular}
\tablefoot{The random errors of the fluxes are less than 1\%. A fiducial flux calibration error of 10\% is taken as a more realistic estimate of the uncertainty.} 
\end{table}

\section{Mass models}
\label{sec: mass models}

\subsection{Bayesian rotation-curve fitting}

In this section, we build a set of mass models with different combinations of mass components (gas, star, dark matter halo). The model circular velocity ($V_{\rm mod}$), therefore, is determined by several free parameters $\vec{p}$ depending on which mass components are included. To determine the parameter values and uncertainties, we use a Markov-Chain-Monte-Carlo (MCMC) method to sample the posterior probabilities of the free parameters (see Appendix~\ref{App: MCMC} for details).

In Bayesian inference, the posterior probability distribution of the free parameters is the product of the likelihood function (based on new observations) and their priors (based on previous knowledge or assumptions). We define the likelihood function as $\mathcal{L}=\exp(-0.5\chi^{2})$ with
\begin{equation}
\chi^{2}=\sum_{k=1}^{N}{\frac{[V_{\rm c}-V_{\rm mod}(\vec{p})]^{2}}{\delta_{V_{\rm c}}^{2}}},
\end{equation}
where $V_{\rm c}$ is the observed circular velocity at the $k-$th radius $R_{\rm k}$ and $\delta_{V_{\rm c}}$ is the associated uncertainty.

Apart from the free parameters in $V_{\rm mod}$, the disk inclination $i$ is treated as a nuisance parameter. We impose a Gaussian prior on $i$ centered at $i_0=53^{\circ}$ and with a standard deviation of $5^{\circ}$ to account for the observational uncertainties (see Section~\ref{sec: disk geometry} and Table~\ref{table: disk geometry}). When sampling in the parameter space of $i$, $V_{\rm c}$ and $\delta_{V_{\rm c}}$ change by a factor of $\sin(i_{0})/\sin(i)$ accordingly.

In the following sections, we explore different mass models and clarify priors on the related free parameters. We start with partial mass models with a limited amount of baryonic components; these models are probably unphysical but are useful to set hard upper limits on gas and stellar masses. Next, we build complete mass models, but warn that the masses of the different components are often degenerate. The best-fit models are shown in Fig.~\ref{fig: partial mass models} \&~\ref{fig: complete mass models} and the MCMC corner plots are shown in Fig.~\ref{fig: Corner plots - partial model} \&~\ref{fig: Corner plots - complete model}. Median values and associated uncertainties of parameters of each model are presented in Table~\ref{table: mass models}.

\subsection{Partial mass models}

\subsubsection{Gas only}
\label{sec: max gas}

To set a hard upper limit on the gas mass, we start with a minimalist mass model where the gas disk is the only dynamically important component. This mass model is probably unphysical; as we will show, indeed, it cannot reproduce the observed rotation curve. 

The gas gravitational contribution ($V_{\rm gas}$) is calculated by numerically solving the Poisson's equation for a finite-thickness disk with a density profile $\rho(R,z)=\Sigma(R)\xi(z)$, where $\Sigma(R)$ is the radial surface density profile and $\xi(z)$ is the vertical profile. To this aim, we use the \texttt{vcdisk} package\footnote{https://github.com/lposti/vcdisk}. For $\Sigma(R)$, we take the \sersic\ profile fitted to the \ci\,(2-1) surface brightness profile in Section~\ref{sec: densityprofile}. For $\xi(z)$, we assume an exponential distribution with a constant scale height of 300 pc.

For practical reasons, we calculate $V_{\rm gas}$ for a normalization mass ($M_{0}$) defined for $R\rightarrow\infty$ and we introduce a dimensionless scaling factor $\Upsilon_{\rm gas}=M_{\rm gas}/M_{0}$ of the order of unity, where $M_{\rm gas}$ is the actual gas mass. Therefore, we have $V_{\rm mod}^{2}=\Upsilon_{\rm gas}V_{\rm gas}^{2}$. For numerical convenience, we take $M_{0}=10^{11}$ \Msun\ and apply hard boundaries on $\log(\Upsilon_{\rm gas}) \in (-2, 2)$. Therefore, $M_{\rm gas}$ has a uniform ``uninformative'' prior within $(10^{9}, 10^{13})$ \Msun.

The best-fit model gives $M_{\rm gas}=8.6\times10^{10}$ \Msun. This value, as we will discuss in Section~\ref{sec: discuss gas mass}, is comparable to the molecular gas mass inferred from the CO $J=4-3$ (hereafter CO\,(4-3)) flux but is three times smaller than those inferred from \ci\ and dust emissions \citep{2024arXiv241104290H}.

The left panel of Fig.~\ref{fig: partial mass models} shows that it is impossible to reproduce the inner parts of the rotation curve using only the gas disk component. The high rotation velocities in the innermost two rings require the existence of a central mass concentration, such as a stellar spheroid and/or a supermassive black hole.

\subsubsection{Stars only}
\label{sec: max star}

We now consider a mass model where $V_{\rm mod}$ is fully determined by the stellar component while the gas contribution is neglected. Since \pks\ is very bright in \ci, this model corresponds to a scenario where the \ci-to-$M_{\rm gas}$ conversion factor is extremely small (see discussions in Section~\ref{sec: discuss gas mass}), so that the gravitational contribution from gas is much smaller than that from stars. This model is probably unrealistic but is useful for setting hard upper limits on the stellar mass.

Given the lack of high-resolution optical/NIR imaging for \pks, we cannot directly compute $V_\star$ using the observed stellar surface brightness profile (as for $V_{\rm gas}$). Therefore, we adopt a sensible parametric function for the stellar mass distribution by assuming a spherical stellar component described by a \sersic\ profile. The stellar gravitational contribution at radius $R$ is then given by
\begin{equation}
    V_\star(R) = \sqrt{ \dfrac{G_{\rm N} M_\star}{R} \dfrac{\gamma(n_\star(3-p), b(R/R_{\rm e,\star})^{1/n})}{\Gamma(n_\star(3-p))} },
\end{equation}
where the fitting parameters are the stellar mass $M_\star$, the stellar half-mass radius $R_{\rm e,\star}$, and the \sersic\ index $n_\star$. The parameters $p$ and $b$ are functions of $n_\star$. The incomplete and complete gamma functions are denoted as $\gamma$ and $\Gamma$, respectively \citep[see][]{Terzic2005}. Similarly to Section~\ref{sec: max gas}, we calculate $V_\star$ for a normalizing mass $M_{0}=10^{11}$ \Msun\ and introduce a dimensionless parameter $\Upsilon_\star = M_\star/M_0$ so that $V_{\rm mod}^2 = \Upsilon_\star V_\star^2$. We apply uniform priors on $n_\star \in (0.5, 10)$, $\log(\Upsilon_\star) \in (-2, 2)$, and $R_{\rm e,\star} \in (0.1, 5)$ kpc.

\begin{figure*}
  \resizebox{\hsize}{!}
  {
  \includegraphics{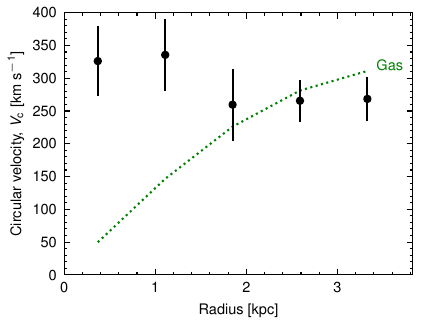}
  \includegraphics{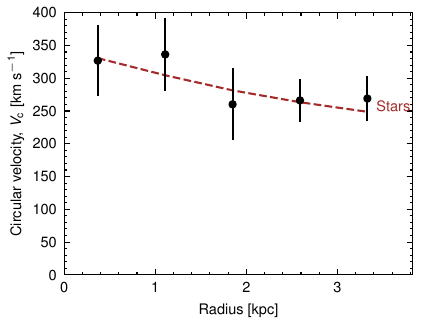}
  }
\caption{Partial mass models: gas only (left panel) and stars only (right panel). In both panels, the black dots with errorbars show the observed circular velocities. The gravitational contributions from gas and stars are shown with a green dotted line and a brown dashed line, respectively.}
\label{fig: partial mass models}
\end{figure*}

The right panel of Fig.~\ref{fig: partial mass models} shows that this single-component model can fit the observed rotation curve. The best-fit $n_\star=5.7$ reconfirms that the stellar mass distribution should be centrally concentrated but $R_{\rm e,\star}$ is unconstrained (see the corner plot in Fig.~\ref{fig: Corner plots - partial model}). Given that the model neglects gas and DM contributions, the best-fit stellar mass ($\sim1.1 \times 10^{11}$ \Msun) represents a hard upper limit on the actual stellar mass of the galaxy. This value is not sensitive to $R_{\rm e,\star}$, as is shown in Fig.~\ref{fig: Corner plots - partial model}, and is about a factor of three smaller than the value estimated from SED fitting \citep[$3\times10^{11}$ \Msun,]{2010ApJ...725...36D}. We will discuss possible reasons for this discrepancy in Section~\ref{sec: discuss stellar mass}.

We have also explored a mass model (Fig.~\ref{fig: MaxDisk}) where the stellar component is given by the sum of an exponential disk and a De Vaucouleurs' bulge (with $n_{\star,bul}=4$). This multi-component mass model has four strongly degenerate parameters: the stellar masses and the effective radii of each component. Since our main aim is to obtain an upper limit on the total stellar mass, the effective radius of the disk was fixed to be equal to that of the dust component $R_{\rm e, dust}$ (Fig.\,\ref{fig: densityprofile}), while that of the bulge was fixed to $0.1\times R_{\rm e, dust}$. This multi-component model also gives a good fit to the rotation curve and returns a total stellar mass (bulge plus disk) of $\sim8\times10^{10}\,M_{\odot}$. This mass is slightly smaller than the one from the single-component spherical model because of the well-known fact that a highly flattened mass distribution implies higher circular velocities than the equivalent spherical mass distribution \citep[e.g.,][]{Lelli2023}. The bulge-to-disk ratio is $\sim0.9$, but this value is highly uncertain and depends on the adopted effective radii. Future high-resolution NIR images are needed to better constrain the stellar mass distribution.

\begin{figure*}
  \resizebox{\hsize}{!}
  {
  \includegraphics{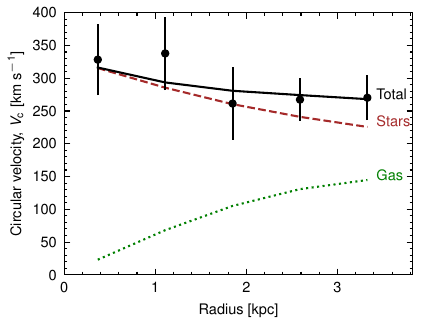}
  \includegraphics{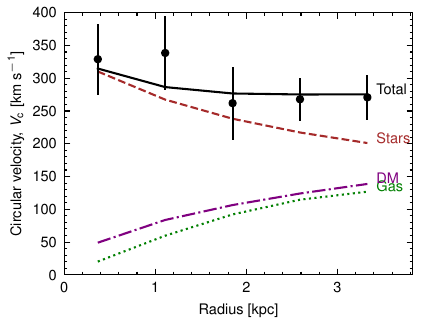}
  }
\caption{Complete mass models: baryons only (left panel) and baryons+DM (right panel). In both panels, the black dots with errorbars show the observed circular velocities, while the black line shows the best-fit mass model. The gravitational contribution from gas, stars, and DM are shown, respectively, with a green dotted line, a brown dashed line, and a purple dash-dotted line.} 
\label{fig: complete mass models}
\end{figure*}

\begin{table*}
\caption{Parameters of the mass models described in Section~\ref{sec: mass models}.}
\label{table: mass models}      
\centering  
\footnotesize
\renewcommand{\arraystretch}{1.2}
\begin{tabular}{ccccccccc}    
\hline\hline
Model & $i$ & $M_{\rm gas}$ & $M_\star$ & $M_{\rm bar}$ & $n_\star$ & $R_{\rm e,\star}$ & $M_{200}$ & $C_{200}$  \\
\cline{3-5}
 & (deg.) & \multicolumn{3}{c}{($10^{10}\,M_{\odot}$)} & & (kpc) & ($10^{12}\,M_{\odot})$ & \\
\hline
Gas only & $53.0_{-4.9}^{+4.8}$ & $8.5_{-1.6}^{+1.7}$ & ...                  & $8.5_{-1.6}^{+1.7}$ & ...      & ...      & ...      & ...      \\
Stars only & $53.0_{-4.9}^{+4.8}$ & ...                 & $10.5_{-4.2}^{+3.4}$ & $10.5_{-4.2}^{+3.4}$     & $5.6_{-2.0}^{+2.3}$ & $3.0_{-1.5}^{+1.4}$ & ...      & ...      \\
Baryons only & $52.5_{-4.9}^{+4.9}$ & $1.8_{-1.8}^{+1.3}$ & $6.9_{-4.6}^{+3.6}$  & $8.7_{-4.9}^{+3.8}$  & $5.0_{-1.5}^{+1.6}$ & $1.8_{-1.2}^{+1.9}$ & ...      & ...      \\
Baryons + DM & $52.4_{-4.9}^{+4.9}$ & $1.4_{-1.4}^{+1.1}$ & $4.7_{-2.8}^{+2.5}$  & $6.1_{-3.2}^{+2.7}$  & $5.1_{-1.7}^{+1.7}$ & $1.3_{-0.8}^{+1.7}$ & $2.2_{-1.3}^{+1.4}$ & $3.6_{-0.9}^{+0.9}$ \\
\hline                  
\end{tabular}
\tablefoot{The values show medians (50\% quantile) with 16\% and 84\% quantiles in sub- and super-scripts, respectively.}
\end{table*}

\subsection{Complete mass models}

\subsubsection{Baryons only}
\label{sec: all baryons}

Compared to the single-component models, a more complete model includes the gravitational contributions of both gas and stars. In this case, $V_{\rm mod}^2=\Upsilon_{\rm gas} V_{\rm gas}^2 + \Upsilon_\star V_\star^2$.

To alleviate the degeneracy among the parameters, we add the following physically-motivated priors:
\begin{enumerate}
    \item A log-normal prior on $\Upsilon_{\rm gas}$. Based on the CO\,(4-3) flux, the CO line ratio $r_{41}=0.46$, and a CO conversion factor $\alpha_{\rm CO}=0.8$, \citet{2024arXiv241104290H} estimate that the molecular gas mass of \pks\ is $7.3\pm0.9\times10^{10}$ \Msun. Considering the uncertainties of the assumed CO conversion factor and the CO line ratio, we center the prior of $\log(\Upsilon_{\rm gas})$ at $-0.14$ with a standard deviation of 0.7 (a factor of 5 for $\Upsilon_{\rm gas}$).  \item A Gaussian prior on $n_\star$. In Section~\ref{sec: max star}, we have demonstrated that the high circular velocities at small radii require the stellar profile to be centrally concentrated. Therefore, we center the prior at $n_\star=4$ with a standard deviation of 2, given that the vast majority of stellar spheroids have \sersic\ indexes between $\sim$2 (such as pseudo-bulges) and $\sim$6 \citep[such as compact ellipticals,][]{2020A&A...644A.117L}.
\end{enumerate}

The left panel of Fig.~\ref{fig: complete mass models} combines the gravitational contribution of both gas and stars. The $M_{\rm gas}$ and $M_\star$ values in this model are $1.8\times10^{10}$ \Msun\ and $7.0\times10^{10}$ \Msun, respectively. As expected, both masses decrease with respect to those in the partial models. The stellar component dominates the total gravitational contribution otherwise the circular velocities at the innermost radii cannot be recovered.

\subsubsection{Baryons plus dark matter}

In Section~\ref{sec: all baryons}, we have shown that the rotation curve of \pks\ is well fitted by a baryon-only model with sensible baryonic masses. Therefore, it is immediately clear that the DM contribution is unconstrained due to the disk-halo degeneracy \citep{1985ApJ...295..305V, Lelli2023}. Nevertheless, we tentatively add a Navarro-Frenk-White (NFW) dark matter halo with constraints based on $\Lambda$CDM cosmology. In this way, we can examine whether the observed rotation curve is consistent with the expectations from the $\Lambda$CDM cosmology.

The NFW-profile is parameterized by the halo concentration $C_{\rm 200}$ and the halo mass $M_{\rm 200}$ (or equivalently the halo velocity $V_{\rm 200}$). Including the contributions of both baryons and the NFW DM halo, $V_{\rm mod}$ is thus given by 
\begin{equation}
V_{\rm mod}^{2}=\Upsilon_{\star}V_{\star}^2 + \Upsilon_{\rm gas}V_{\rm gas}^{2}+V_{\rm NFW}^{2}(M_{200}, C_{200}),
\end{equation}
where $V_{\rm NFW}$ is the circular velocity of the NFW halo \citep[Eq.~10 in][]{2020ApJS..247...31L}. In addition to the baryonic priors described in Section~\ref{sec: all baryons}, we use two $\Lambda$CDM scaling relations as DM priors:
\begin{enumerate}

    \item A log-normal prior on $M_{200}$. \citet{2019MNRAS.486.5468L} determined the stellar-to-halo mass relations in different redshift bins using a parametric abundance matching technique. Here, we relate the mean $M_{200}$ to $M_\star$ through their Eq.~1 with their best-fit parameters at $z = [2,2.5]$. A conservative scatter in $\log(M_{200})$ of 0.2 is adopted.
    
    \item A log-normal prior on $C_{200}$. \citet{2014MNRAS.441.3359D} fitted the $M_{200}-C_{200}$ relation from N-body cosmological simulations. The mean $C_{200}$ is related to $M_{200}$ through their Eq.~7 with redshift-dependent parameters given by their Eq.~10 and Eq.~11. We adopt a scatter of 0.11 in $\log(C_{200})$.
    
\end{enumerate}

Compared to the baryons-only model, including a DM halo decreases the gas mass and stellar mass, but these different estimates are all consistent within uncertainties. Even though the DM contribution is not constrained, the rotation curve of \pks\ is consistent with the expectations from the $\Lambda$CDM cosmology.

\section{Discussion}
\label{sec: discussion}

\pks\ is about $10-100$ times more luminous in the \ci\ line than the majority of high-$z$ radio galaxies \citep[HzRG,][]{2023MNRAS.525.5831K}, enabling detailed studies of its gas distribution and kinematics. On the other hand, the SFR of \pks\ \citep[$1020_{-170}^{+190}$ \Msun\ yr$^{-1}$,][]{2019A&A...621A..27F} indicates that its ISM condition should be extreme (e.g., strong UV field, cosmic ray intensity, gas turbulence, and high gas density and temperature), which could complicate the abundances of molecular gas tracers and the excitation of molecular lines.

\subsection{Circular and non-circular motions}
\label{sec: discuss non-circular motions}

In Section~\ref{sec: rotation curve fitting}, we show that \pks\ has a regular rotating disk, with $V_{\rm rot}/\sigma_{\rm v}=6\pm3$. This value is larger than what is predicted by the disk-instability model from \citet{2015ApJ...799..209W} at the redshift of \pks, but is consistent with recent ALMA observations in a significant sample of high-$z$ star-forming galaxies \citep{2023A&A...672A.106L,2023A&A...679A.129R,2024A&A...689A.273R}. Given that \pks\ is an AGN-host starburst with an extreme SFR of $1020_{-170}^{+190}$ \Msun\ yr$^{-1}$ \citep{2019A&A...621A..27F}, it is surprising that its gas disk is still dynamically cold.

In addition to the overall regular rotation of the gas disk, there are clear signatures of non-circular motions, i.e, the SW-tail, the E-tail, and the anomalous structure (Section~\ref{sec: non-circular}). The gas tails may be remnants of a past major merger event, which could have triggered a gas inflow (possibly related to the anomalous kinematic structure near the center) and therefore the high star-formation rate and radio-loud AGN activity of the galaxy. Alternatively, the two gas tails may be spiral arms in a more extended gas disk, while the kinematically anomalous component may be something unrelated, such as a gas outflow. Future images from the Hubble Space Telescope (HST) or the James Webb Space Telescope (JWST) are key to elucidating their origins.

Considering the fraction of non-circular motions in the total flux of \pks\ as well as the gas mass obtained using the ``gas-only'' mass model, we obtain hard upper limits on the gas mass of the non-circular structures, which is about $1.0\times10^{10}$ \Msun. Here we assume that the flux-to-mass conversion factors are the same in the rotating disk and the non-circular structures. If we take the gas mass from the ``Baryons+DM'' mass model, the mass of the non-circular motions decreases to $1.7\times10^9$ \Msun. In both cases, the molecular gas involved in the non-circular components is a minor fraction (12\%) of the total gas mass that resides in the rotating disk.

\subsection{Discrepancies in different mass estimates} 

Mass models fitted to the observed \ci\ rotation curve allow us to obtain dynamical upper limits on the gas and stellar masses of this galaxy. In the following, we compare our mass measurements with those from independent methods, finding some puzzling discrepancies.

\subsubsection{Discrepancies in gas masses and conversion factors}
\label{sec: discuss gas mass}

To estimate the total molecular gas mass ($M_{\rm mol}$) of a galaxy, one usually measure the line luminosity of an H$_{2}$-tracer and adopt a luminosity-to-mass conversion factor. For example, CO lines have been widely used \citep{2013ARA&A..51..105C}. The CO-to-H$_{2}$ conversion factor $\alpha_{\rm CO}$ is typically defined as the ratio between $M_{\rm mol}$ and the luminosity of the CO\,(1-0) line, $L_{\rm CO\,(1-0)}^{'}$ \citep{2013ARA&A..51..207B}. This conversion factor must be calibrated with an independent measurement of $M_{\rm mol}$, such as the one derived with dynamical methods. By doing so, the underlying assumption is that molecular gas dominates the total gas mass ($M_{\rm gas}$) in the inner galaxy regions.

In the case of \pks, using the CO\,(4-3) luminosity $L_{\rm CO\,(4-3)}^{'}=(4.2\pm0.5)\times10^{10}$ K \kms\ pc$^{2}$ \citep{2024arXiv241104290H}, a typical CO line ratio of $r_{41}\equiv L_{\rm CO\,(4-3)}^{'}/L_{\rm CO\,(1-0)}^{'}=0.5$ \citep{2013ARA&A..51..105C}, and the upper-limit on $M_{\rm gas}$ from the gas-only mass model (Section~\ref{sec: max gas}), we get an upper limit on $\alpha_{\rm CO}<1.0\,(r_{41}/0.5)$ \Msun\ (K km s$^{-1}$ pc$^{2}$)$^{-1}$. This is similar to what is commonly used for starbursts \citep[$\sim$0.8,][]{2013ARA&A..51..207B,2013ARA&A..51..105C}, but we stress that it is a very hard upper limit because it neglects contributions from stars and DM in the mass model. If we instead consider the gas mass from the complete mass model with baryons plus DM, we find $\alpha_{\rm CO} = 0.17\,(r_{41}/0.5)$ \Msun\ (K km s$^{-1}$ pc$^{2}$)$^{-1}$.

Using the \ci\,(1-0) luminosity $L_{\rm [C\,{\sc I}]}^{'}=(3.12\pm0.67)\times10^{10}$ K \kms\ pc$^{2}$ \citep{2024arXiv241104290H} and the upper-limit of $M_{\rm gas}$ from the gas-only model, we get an upper limit on the \ci-to-$M_{\rm gas}$ conversion factor $\alpha_{\rm [C\,{\sc I}]}\equiv M_{\rm gas}/L_{\rm [C\,{\sc I}]}^{'}<2.8\pm0.8$ \Msun\ (K km s$^{-1}$ pc$^{2}$)$^{-1}$. This value is about 1/7 of the mean value in high-$z$ metal-rich galaxies \citep[$\sim$20,][]{2022MNRAS.517..962D}. If we consider the gas mass from the complete mass model with baryons plus DM, the inferred value of $\alpha_{\rm [C\,{\sc I}]}$ goes down to 0.4 \Msun\ (K km s$^{-1}$ pc$^{2}$)$^{-1}$, which is 50 times lower than the usual value. 

One possibility is that \pks\ has a \ci/H$_{2}$ abundance ratio at least seven times the value taken for local ultra-luminous infrared galaxies \citep[$\sim3\times10^{-5}$,][]{2004ApJ...615L..29P}. If \pks\ is the progenitor of a local early-type galaxy (ETG), we may indeed expect that its star-forming gas is already significantly enriched \citep[e.g.,][]{2005ApJ...621..673T,2010MNRAS.404.1775T}. Moreover, considering the intense star formation and AGN activity in \pks, the \ci/H$_{2}$ ratio can also be enhanced by the dissociating far-UV photons and cosmic rays \citep{2024MNRAS.527.8886B}.

Another possibility is that the radiative transfer of \ci\ lines is complicated by the intense star formation and AGN activity, leading to enhanced \ci\ emission and the failing of the usual conversion factor. Moreover, the \ci\,(1-0) flux is very uncertain and there may be spatial variations of the \ci\ line ratio that we cannot probe with the current data.

\subsubsection{Discrepancies in stellar masses}
\label{sec: discuss stellar mass}

The upper limit on $M_\star$ given by the star-only mass model is about a factor of $\sim3$ smaller than the value estimated from fitting the spectral energy distribution (SED) with stellar population models \citep{2010ApJ...725...36D}. The discrepancy increases up to a factor of $\sim6$ if the gravitational contributions of gas and DM are included in the mass models. The discrepancy is quite serious, so we discuss three possibilities to explain it: (1) the SED fitting overestimates the stellar mass; (2) the rotating disk is not in full equilibrium, so the circular velocities underestimate the dynamical mass; and (3) we are observing two different galaxies along the line of sight.

(1) Regarding the SED fitting, the stellar mass comes from \citet{2010ApJ...725...36D}, where 70 HzRGs were studied based on Spitzer photometry. The SED fitting assumes an elliptical galaxy template for the stellar component, which may not be ideal for \pks\ given its high SFR \citep{2019A&A...621A..27F}. Two or three black body functions are assumed for dust emissions. The inherent uncertainty of stellar mass derived from SED fitting \citep{Seymour2007} is smaller than its difference from the stellar mass derived using dynamic mass models. To further investigate the issue, we have constructed a new SED with 22 photometric points from the rest-frame optical to the radio, using data from Gemini Flamingos-2, VLT Infrared Spectrometer And Array Camera (ISAAC), Spitzer/MIPS and IRAC, Herschel/SPIRE and PACS, ALMA band 4 and band 6, and ATCA. Preliminary SED fittings with Code Investigating GALaxy Emission \citep[CIGALE,][]{Boquien2019} show that the best-fit stellar mass can range from $0.4\times10^{11}$ to $1.2\times10^{11}$ M$_\odot$ depending on the chosen AGN model, so it may be consistent with the dynamically-inferred value of $M_\star$. The SED fittings, however, are not fully satisfactory, especially in the AGN-dominated far-infrared portion of the spectrum, so we will investigate this issue in more detail in a future paper, in which we will test different SED fitting codes and AGN models.

(2) Regarding the dynamical equilibrium, the \ci\,(2-1) velocity field is relatively symmetric and shows regular rotation (Fig.\,\ref{fig: moment-maps}), which is usually interpreted as the cold gas being in equilibrium with the gravitational potential. Our 3D kinematic modeling, however, reveals an anomalous kinematic structure in the approaching side of the disk (Fig.\,\ref{fig: ci-pv}) and two extended gaseous tails. If \pks\ has indeed undergone a recent major merger, the inner disk may not have had enough time to relax with the overall gravitational potential, so that the dynamical mass is potentially underestimated \citep{2015A&A...584A.113L}. 

Using the outermost measured point of the rotation curve, we estimate the orbital time of \pks, which is $t_{\rm orb}\sim80$ Myr. This is larger than the time from the two recent bursts of star formation: 6 Myr and $>20$ Myr \citep{2019A&A...624A..81M}. If the two star formation bursts are driven by a major merger, it is therefore possible that the gas disk of \pks\ is not relaxed because it did not have enough time to complete several rotations since the time of the latest starburst. Custom-built hydrodynamical simulations are needed to investigate whether such a merger event could be strong enough to drive the rotating disk out of dynamical equilibrium, leading to a systematic underestimate of the dynamical mass.

(3) Regarding the third possibility, the scenario is that we are seeing two galaxies roughly aligned along the line of sight: a \ci-emitting, gas-rich, star-forming galaxy on the foreground and a \oiii-emitting, gas-poor, AGN-dominated galaxy on the background. As strange as it may sound, there are actually several clues in this direction. 

\begin{figure*}[h!]
  \resizebox{\hsize}{!}
  {
  \includegraphics{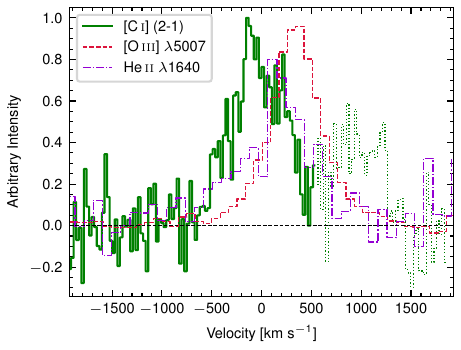}
  \includegraphics{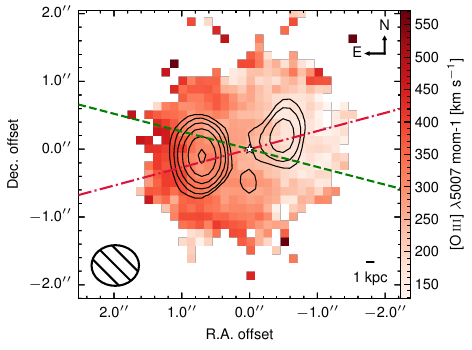}
  }
   \caption{Comparison among ALMA, VLT/SINFONI, and VLT/X-Shooter data \citep[see also][]{2018MNRAS.479.5440L}. Left: The \ci\,(2-1) (green solid) and \oiii\,$\lambda$5007 (red dashed) line profiles are extracted from an aperture of 1-arcsec in diameter centered at the kinematic center. The He\,{\sc ii}\,$\lambda$1640 (violet dash-dotted) line is adapted from \citep{2019A&A...624A..81M}. The velocities of \ci\,(2-1), \oiii\,$\lambda$5007, and He\,{\sc ii}\,$\lambda$1640 are relative to the redshift of \ci\,(2-1) line ($z=2.5706\pm0.0002$). The CO\,(7-6) emission adjacent to \ci\,(2-1) is marked with green dotted line (not shifted in velocity). Right: The \oiii\ velocity field \citep[adapted from][]{2017A&A...599A.123N,2018MNRAS.479.5440L} is overlaid with the radio lobes from ATCA 18-GHz (black contours). The line-of-sight velocities are with respect to the redshift of \ci\,(2-1). The kinematic major axes of the \ci\,(2-1) and \oiii\,$\lambda5007$ lines are shown by the green dashed line and the red dash-dotted line, respectively. The FWHM of the SINFONI point-spread function (0.7 arcsec $\times$ 0.6 arcsec) is shown in the bottom left corner. The axis coordinates are relative to the kinematic center (white star).}
   \label{fig: spec-compare}
\end{figure*}

First, the \oiii\,$\lambda$5007 emission is systematically redshifted with respect to the \ci\,(2-1) emission (Fig.~\ref{fig: spec-compare}). The redshifts of \ci\,(2-1) and \oiii\,$\lambda5007$ lines are $2.5706\pm0.0002$ and $2.5745\pm0.0001$ \citep{2017A&A...599A.123N}, respectively. Their redshift difference corresponds to either a velocity difference of $\sim350$ \kms\ with respect to the \ci\,(2-1) rest frame, or a co-moving distance difference of $\sim4.5$ Mpc if we consider the \ci\ and \oiii\ redshifts as distinct reference frames. Given the cosmological scale-factor of 0.28 at $z=2.57$, the physical distance between the \ci\ and \oiii\ emitters would be of 1.3 Mpc, so the two putative galaxies would probably be unbound.

Second, the \oiii\,$\lambda$5007 kinematic major axis is offset by $\sim30^{\circ}$ with respect to the \ci\ disk major axis (see Fig.\,\ref{fig: spec-compare}). This fact was already noticed by \citet[][see their Fig. 1]{2018MNRAS.479.5440L}, who interpreted the \oiii\ emission as coming from the redshifted, far-side of an ionized gas outflow, given that the \oiii\ kinematic major axis is well aligned with the AGN-driven radio lobes. 

Discrepant redshifts from several different lines were also found in \citet{2019A&A...624A..81M} using rest-frame UV absorption lines from VLT/X-Shooter observations. This two-galaxies scenario is similar to the configuration of the Dragonfly Galaxy \citep{2023ApJ...951...73L}, where two galaxies (though both gas-rich) are merging, while one of them hosts an AGN and two radio lobes. To test this scenario, we need high-resolution images from HST or JWST to possibly discern two separate stellar components.

\subsection{The disk-halo degeneracy}

In the previous sections, we discussed discrepancies between stellar and gas masses from ``photometric'' and ``dynamical'' methods. These discrepancies already emerge when we consider single-component mass models, which provide hard upper limits to the mass of each individual component. Clearly, the discrepancies become even more severe when we consider two-component models (gas and stars) or multi-component models with a DM halo. These facts highlight the severity of the disk-halo degeneracy at high $z$ \citep{Lelli2023}: if we cannot measure with high confidence the stellar and gas masses with ``photometric'' methods, there is little hope to measure the DM content.

The disk-halo degeneracy has been a long-standing issue in building mass models at $z=0$ \citep{1985ApJ...295..305V}. In particular, \citet{1986RSPTA.320..447V} showed that one needs to know the baryonic mass with an accuracy of about 25\% to fully break the degeneracy, even when extended rotation curves from H\,{\sc i} observations are available (see their Fig.~5). For galaxies at cosmic noon, the stellar masses from SED fitting and the gas masses from standard methods (often based on high-$J$ CO lines) are surely more uncertain than 25\%, indicating that major observational and technical endeavours are needed to address the crucial question of the DM content of high-$z$ galaxies.

In recent years, several studies reported DM fractions of galaxies at cosmic noon \citep{Price2021, NestorShachar2023, Puglisi2023} and some of them even argued to find evidence for DM cores \citep{Genzel2020, Bouche2022}. These works, however, rarely discuss or investigate the disk-halo degeneracy, possibly indicating some over-confidence in knowing the true baryonic masses of high-$z$ galaxies. At $z\simeq0$, one approach to break the disk-halo degeneracy has been to use NIR surface photometry in combination with dedicated stellar population models \citep{Schombert2014, Schombert2019}. Even so, some systematic uncertainties remain due to the choice of the specific stellar population model and stellar initial mass function, so additional dynamical arguments are used to set the absolute calibration of the stellar mass \citep{McGaugh2015, Lelli2016a, Lelli2016b, Lelli2016c}. At high $z$, the current situation is much more uncertain, but rest-frame NIR imaging with JWST may be a promising route to measure robust stellar masses, while multi-line gas tracer observations may allow to measure robust gas masses, so that the disk-halo degeneracy could be ameliorated, using stringent, physically motivated priors when fitting the rotation curve.

\section{Conclusions}
\label{sec: conclusions}

In this work, we study the gas distribution and dynamics of a radio-loud AGN-host galaxy at $z\simeq2.6$, \pks, using ALMA data of the \ci\,(2-1) line with a superb spatial resolution of 0.18$''$ ($\sim$1.5 kpc). Our results can be summarized as follows:

\begin{enumerate}

\item The \ci\,(2-1) emission forms a dynamically cold, rotation supported disk with $V_{\rm rot}/\sigma_{\rm v}=6\pm3$, confirming the overall picture from low-resolution data \citep{2018MNRAS.479.5440L};

\item We discover two gas tails extending beyond the rotating disk and a kinematically anomalous gas component at $\sim$2 kpc from the galaxy center. These non-circular structures may be related and be due to a past merger event;

\item Our 3D kinematic modeling returns a flat rotation curve at large radii, which implies a total dynamical mass of $\sim$10$^{11}$ \Msun\ within about 3.3 kpc;

\item Mass models with multiple components display a strong disk-halo degeneracy: models with or without a DM halo can explain equally well the observed circular velocities, so the DM content is virtually unconstrained;

\item The dynamical upper limit on $M_\star$ is exceeded by the stellar masses from available SED fitting, while the dynamical upper limit on $M_{\rm gas}$ is exceeded by gas masses from usual recipes. The origin of these discrepancies remain unclear.

\end{enumerate}

High-resolution optical/NIR images, such as those from HST and/or JWST, are needed to probe the stellar mass distribution and break the disk-halo degeneracy, so to measure the actual DM content of high-$z$ galaxies. These images may also help to understand the discrepancies between the different methods for estimating stellar and gas masses at high $z$, which are key aspects to understand the formation and evolution of galaxies.

\begin{acknowledgements}

L. L. and F. L. acknowledge the hospitality of ESO Garching, where most of this work was done. L.L. and Z.Y.Z acknowledge the support from the National Key R\&D Program of China (2023YFA1608204). L.L. and Z.Y.Z acknowledge the support of the National Natural Science Foundation of China (NSFC) under grants 12173016 and 12041305. L.L. and Z.Y.Z acknowledge the science research grants from the China Manned Space Project, CMS-CSST-2021-A08 and CMS-CSST-2021-A07. L.L. and Z.Y.Z acknowledge the Program for Innovative Talents, Entrepreneur in Jiangsu. A. M. acknowledges the support of the Natural Sciences and Engineering Research Council of Canada (NSERC) through grant reference number RGPIN-2021-03046. T.G.B. acknowledges support from the Leading Innovation and Entrepreneurship Team of Zhejiang Province of China (Grant No. 2023R01008). This paper makes use of the following ALMA data: ADS/JAO.ALMA\#2018.1.01669.S. ALMA is a partnership of ESO (representing its member states), NSF (USA) and NINS (Japan), together with NRC (Canada), NSTC and ASIAA (Taiwan), and KASI (Republic of Korea), in cooperation with the Republic of Chile. The Joint ALMA Observatory is operated by ESO, AUI/NRAO and NAOJ.

\end{acknowledgements}

\bibliographystyle{aa} % style aa.bst
\bibliography{mybib.bib} % your references Yourfile.bib


\begin{appendix} %First appendix

\section{Posterior probability distribution from Markov-Chain Monte-Carlo fits}
\label{App: MCMC}

\begin{figure*}[h!]
  \resizebox{\hsize}{!}
  {
    \includegraphics[width=0.5\hsize]{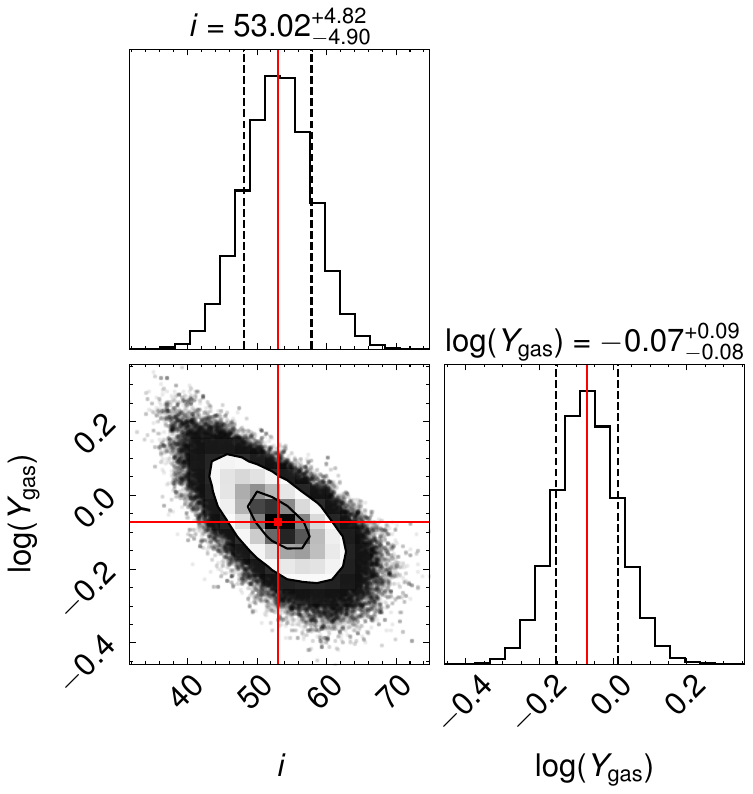}
    \includegraphics[width=0.5\hsize]{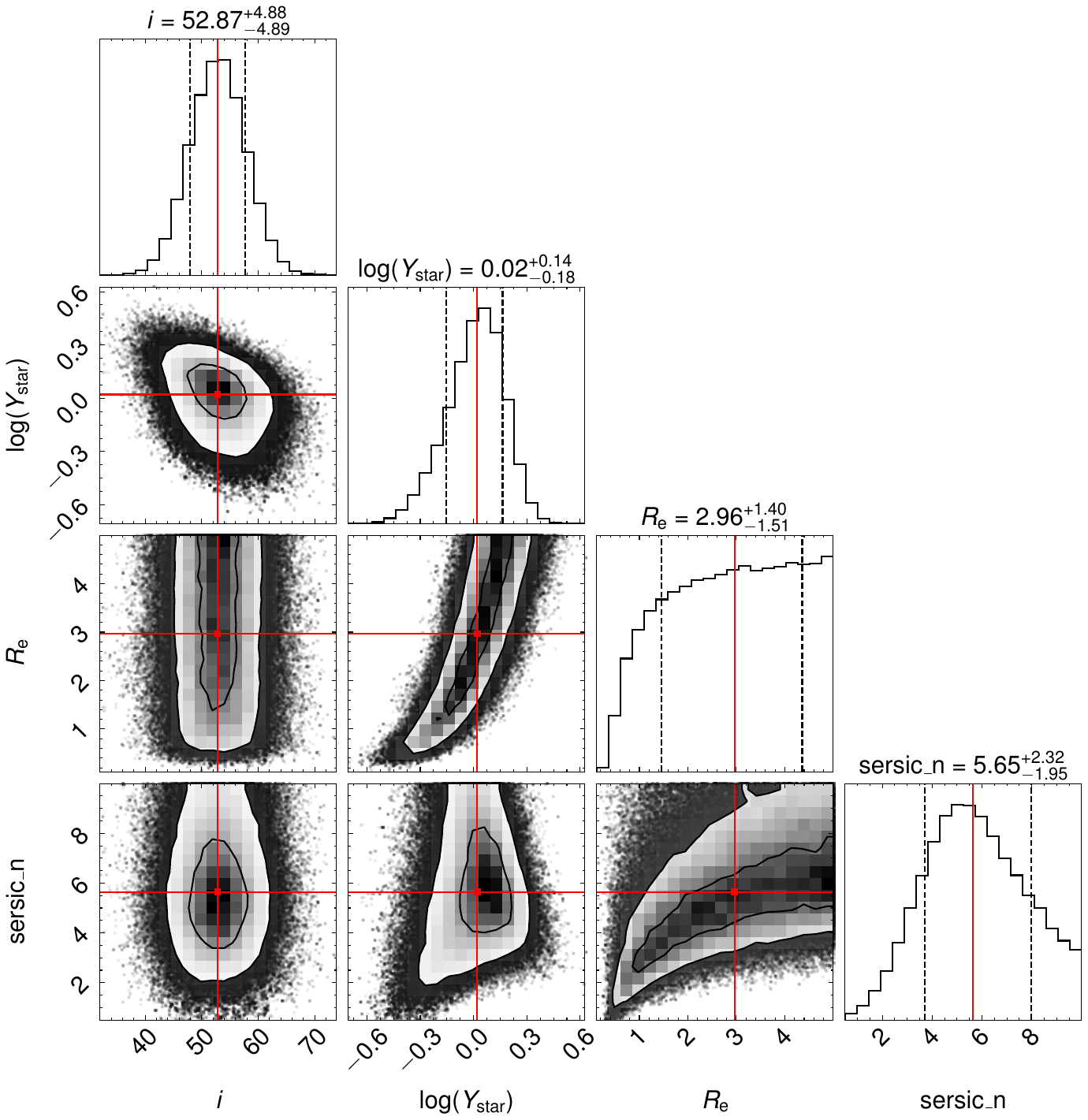}
  }
\caption{Corner plots for partial mass models: gas-only model (left) and stars-only model (right). See Section~\ref{sec: mass models} for details.}
\label{fig: Corner plots - partial model}
\end{figure*}

\begin{figure*}[h!]
  \resizebox{\hsize}{!}
{
\includegraphics[width=0.5\hsize]{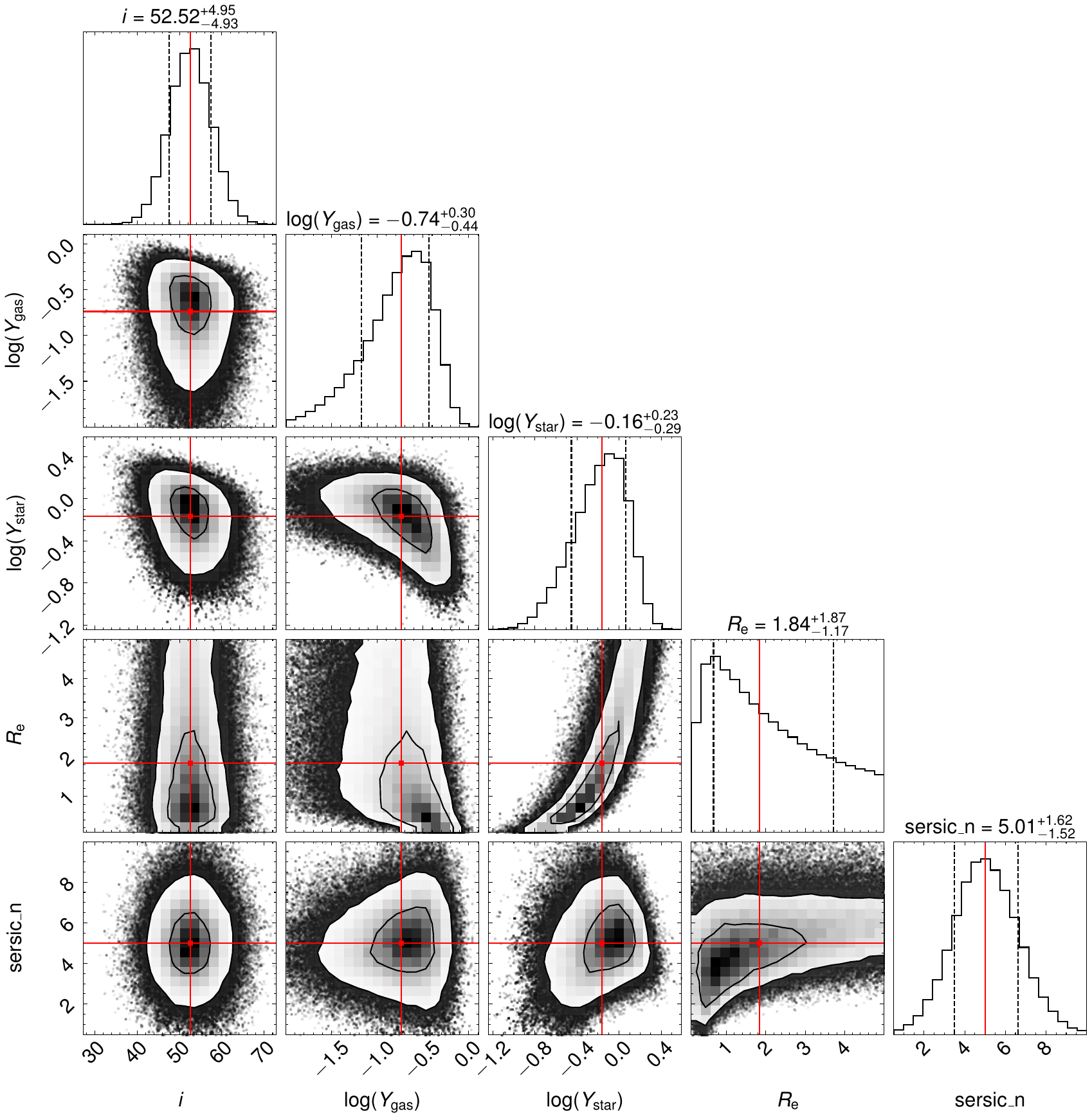}
\includegraphics[width=0.5\hsize]{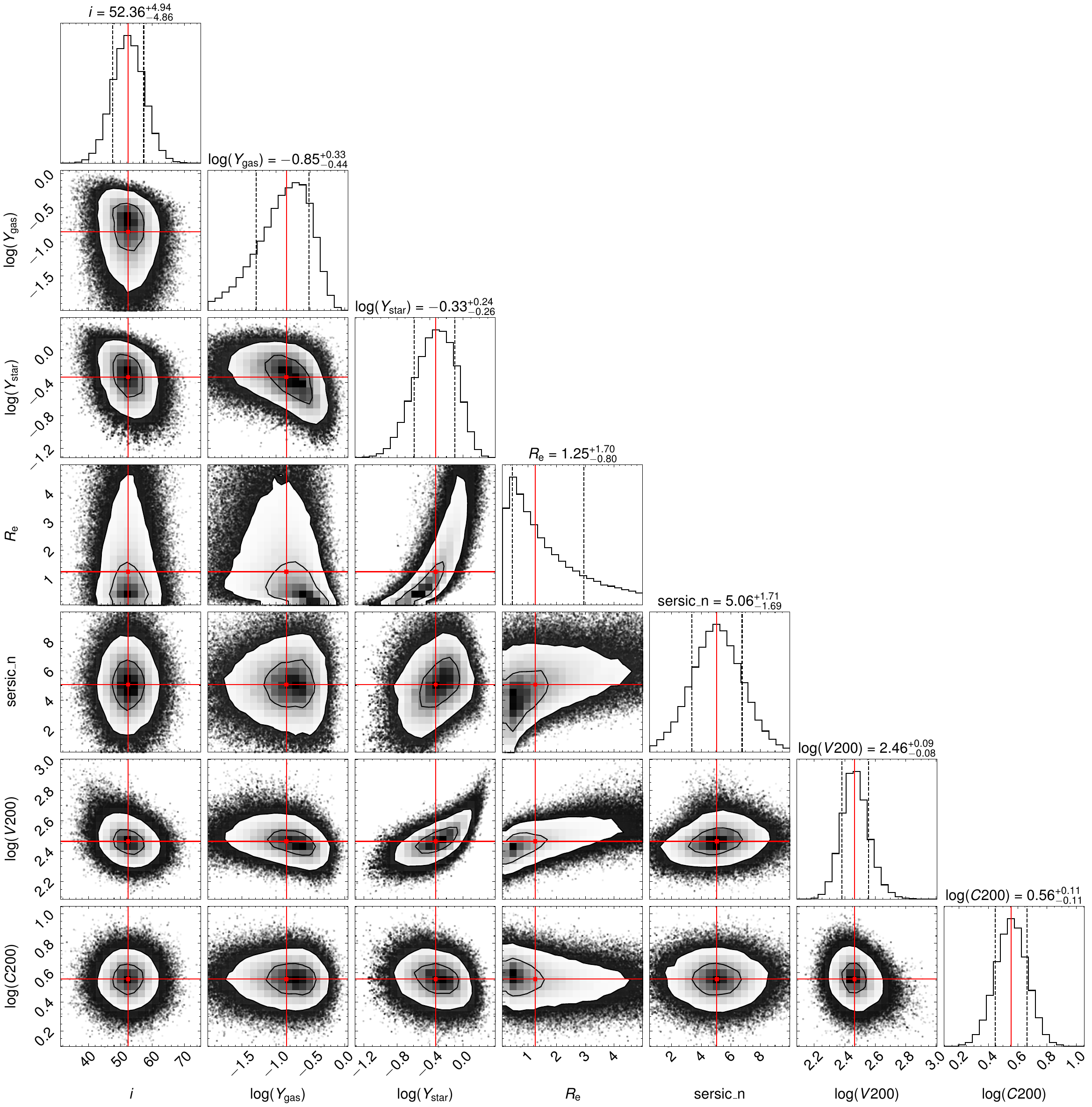}
}
\caption{Corner plots for complete mass models: baryons-only model (left) and baryons-plus-DM model (right). See Section~\ref{sec: mass models} for details.}
\label{fig: Corner plots - complete model}
\end{figure*}

Fig.\,\ref{fig: Corner plots - partial model} and Fig.\,\ref{fig: Corner plots - complete model} show ``corner plots'' from MCMC fits to the rotation curves (see Section~\ref{sec: mass models}). The corner plots are obtained using the \texttt{corner} package \citep{2016JOSS....1...24F}. The various panels of the corner plots show the posterior probability distribution of pairs of the fitting parameters (inner panels) as well as the marginalized 1D probability distribution of each parameter (outer panels). In the inner panels, individual MCMC samples outside the 2$\sigma$ confidence region are shown with black dots, while binned MCMC samples inside the 2$\sigma$ confidence region are shown by a grayscale; the black contours correspond to the 1$\sigma$ and 2$\sigma$ confidence regions. In the outer panels (histograms), red solid lines and dashed black lines correspond to the median and $\pm1\sigma$ values, respectively. The red solid lines continue in the outer panels, hitting the median value of the parameter (red square).

\begin{figure*}[h!]
\centering
  \resizebox{\hsize}{!}
{
\includegraphics[width=0.55\hsize]{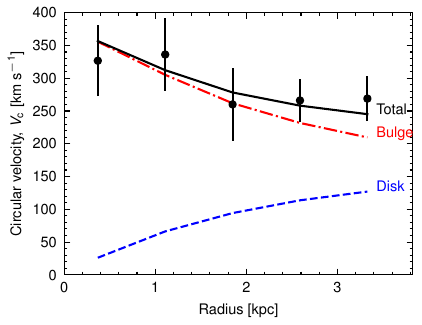}
\includegraphics[width=0.45\hsize]{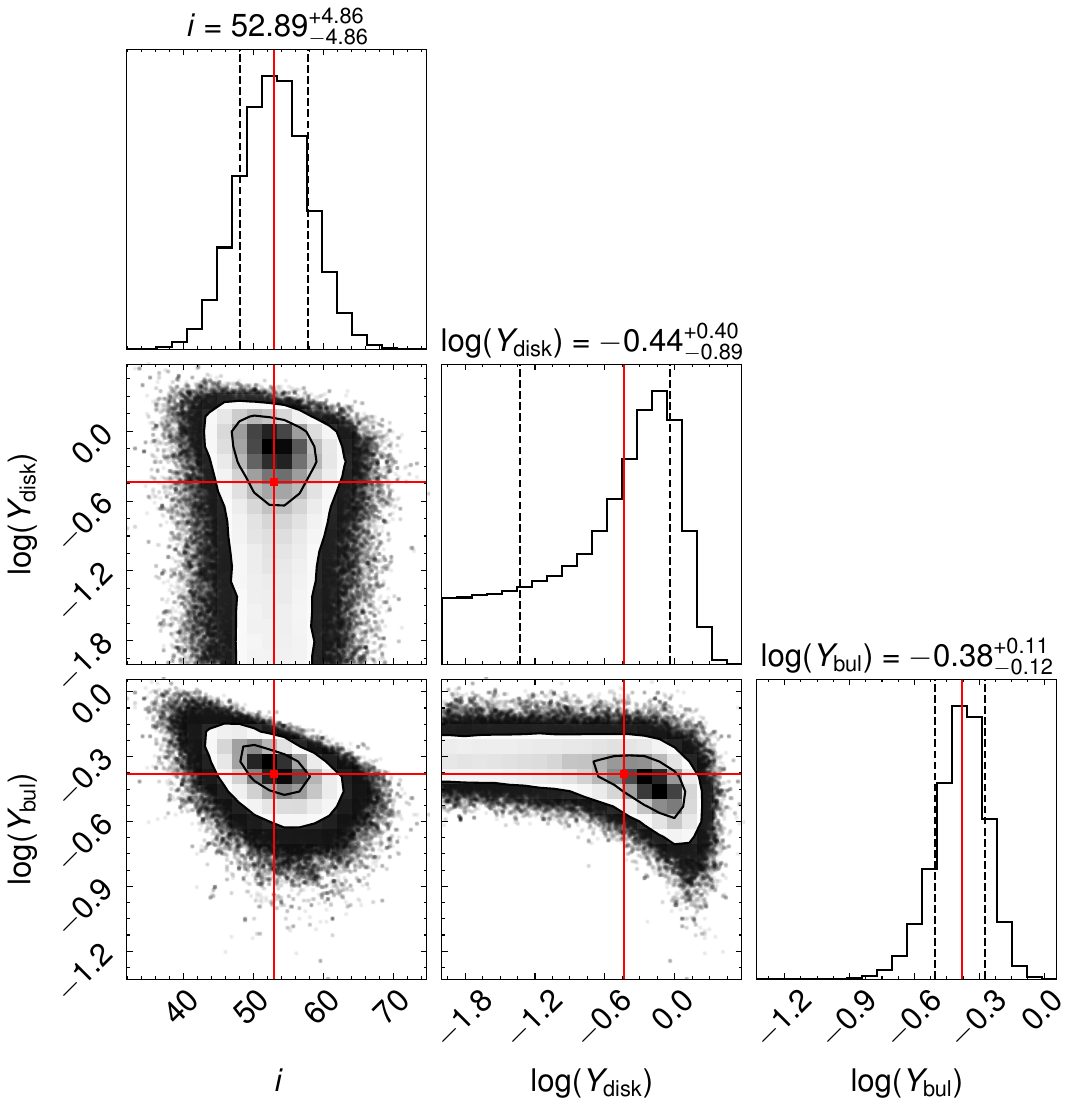}
}
\caption{Left: A mass model where the stellar component is composed of an exponential disk (blue dashed line) and a De Vaucouleurs bulge (red dash-dotted line). The black dots with errorbars show the observed circular velocities, while the black line shows the best-fit mass model. This model gives a total stellar mass (bulge plus disk) of $8.0_{-7.8}^{+3.5}\times10^{10}$ \Msun. Right: Corner plot for the stellar disk + bulge mass model.}
\label{fig: MaxDisk}
\end{figure*}

The four corner plots correspond to the different mass models presented in Section~\ref{sec: mass models}, having an increasing number of mass components and free parameters. In addition, we show in Fig.~\ref{fig: MaxDisk} a mass models where the stellar component is divided up in a thick exponential disk and spherical De Vaucouleurs' bulge. In general, the posterior probability distributions are well-behaved and show clear peaks, indicating that the fitting quantities are well measured. The only exception is represented by the effective radius of the stellar spheroid ($R_{\rm e}$) which is poorly constrained in all models, so it should be interpreted as a fiducial upper limit.

% \FloatBarrier

\end{appendix}

\end{document}